\documentclass[aps,prx,reprint,10pt,superscriptaddress,amssymb,amsfonts,longbibliography,footinbib]{revtex4-1}

\usepackage{amssymb}
\usepackage{amsmath}
\usepackage{commath}
\usepackage{graphicx}
\usepackage{verbatim}
\usepackage{mathrsfs}
\usepackage{color}
\usepackage{xcolor}
\usepackage{floatrow}
\usepackage{physics}
\usepackage{dcolumn}
\usepackage{bm}
\usepackage{MnSymbol}
\usepackage[normalem]{ulem}
\usepackage{mathtools}
\usepackage{wasysym}
\usepackage{ulem}
\usepackage{hyperref}
\usepackage{soul,xcolor}
\usepackage{lineno}
\usepackage{enumitem} 
\usepackage[T1]{fontenc}
\usepackage{lipsum}
\usepackage{amsmath, array} 
\usepackage[normalem]{ulem}

\begin{document}

\title{
Phonon-induced disorder in dynamics of optically pumped metals from\\ non-linear electron-phonon coupling
}
 
\author{John Sous}\email{js5530@columbia.edu}
\affiliation{Department of Physics, Columbia University, New York, New York 10027, USA}

\author{Benedikt Kloss}
\affiliation{Department of Chemistry, Columbia University, New York, New York 10027, USA}

\author{Dante M. Kennes}
\affiliation{Institut f\"{u}r Theorie der Statistischen Physik, RWTH 
Aachen, 52056 Aachen, Germany and JARA - Fundamentals of Future 
Information Technology}

\affiliation{Max Planck Institute for the Structure and Dynamics of 
Matter and Center for Free-Electron Laser Science, 22761 Hamburg, Germany}

\author{David R. Reichman}\email{drr2103@columbia.edu}
\affiliation{Department of Chemistry, Columbia University, New York, New York 10027, USA}

\author{Andrew J. Millis}\email[Correspondence and requests for materials should be addressed to A.~J.~M.: ]{ajm2010@columbia.edu}
\affiliation{Department of Physics, Columbia University, New York, New York 10027, USA}
\affiliation{Center for Computational Quantum Physics, Flatiron Institute, 162 5th Avenue, New York, New York 10010, USA}

\date{\today}

\begin{abstract} 
The non-equilibrium dynamics of matter excited by light may produce electronic phases that do not exist in equilibrium, such as laser-induced high-transition-temperature superconductivity. Here we simulate the dynamics of a metal driven at initial time $t=0$ by a spatially uniform pump that excites dipole-active vibrational modes which couple quadratically to electrons. We study in detail the evolution of electronic and vibrational observables and their coherences.  We provide evidence for enhancement of local electronic correlations, including double occupancy, accompanied by rapid loss of spatial structure, which we interpret as a signature of emergent effective disorder in the dynamics.  This effective disorder, which arises in absence of quenched randomness, dominates the electronic dynamics as the system evolves towards a correlated electron-phonon long-time state, possibly explaining why transient superconductivity is not observed. The pumped electron-phonon systems studied here, which are governed by non-linear coupling, exhibit a much more substantial dynamical response than linearly coupled models relevant in equilibrium, thus presenting a pathway to new modalities for out-of-equilibrium phases. Our results provide a basis within which to understand correlation dynamics in current pump-probe experiments of vibrationally coupled electrons, highlight the importance of the evolution of phase coherence, and demonstrate that pumped electron-phonon systems provide a means of approximately realizing recently proposed scenarios of dynamically induced disorder in translation-invariant systems.
\end{abstract}

\maketitle

\normalsize

Major efforts in condensed-matter physics are currently focused on the means to induce novel phases of matter and harness their properties for practical gain.  For many years such phases were thought to robustly exist only as equilibrium, thermodynamic states.  The potential out-of-equilibrium induction of transient phases, enabled by recent experimental advances in the creation and utilization of tailored time-resolved external fields that can excite specific degrees of freedom, opens a door to new modalities for the realization and control of new electronic states \cite{RMPBasov,RevGiannetti}.

Optical, mode-specific excitation of atomic vibrations \cite{ModeSpecificCavalleri} serves as one broad class of out-of-equilibrium techniques that has been shown experimentally to lead to dramatic modifications in electronic behavior \cite{Merlin,ManganiteCav,WTe}, including the possible induction of a superconducting transition at a critical temperature larger than its equilibrium counterpart in K$_3$C$_{60}$ \cite{K3C60}, YBa$_2$Cu$_3$O$_{6.5}$ \cite{CuprateCav} and organic salts \cite{SaltCav}.  In general, optically accessible phonons are long-wavelength dipole-active modes, which do not couple linearly to the electron density, and therefore non-linearities are expected to govern the dynamics in centrosymmetric systems \cite{NonLinear1,NonLinear2,NonLinear3,NonLinear4Kennes}, stimulating many interesting theoretical proposals \cite{NonLinear4Kennes,KollathSentef,GeorgesFullerides,Demler1,QuadraticSentef,Demler2,EcksteinWerner,EcksteinNonLinear}. One particular mechanism \cite{NonLinear4Kennes} is based on the observation that since direct, local interaction between electrons and photo-excited phonons must depart from that of conventional linear (Holstein \cite{Holstein} and Fr{\"o}hlich \cite{Froh1,Froh2}) models, one must consider a quadratic coupling of driven phonons to the electron density. An approximate analysis of such a model was presented previously~\cite{NonLinear4Kennes} (see Supplementary Information). Here, we use \emph{exact} numerical methods and an effective theory based on a low-order expansion in the electron-phonon coupling to unravel the emergent electronic behavior in this driven, non-equilibrium system. Combining a tensor-network approach for time evolution of an infinite one-dimensional system on short timescales with propagation to long times using direct Krylov subspace methods for finite-size systems and analytical arguments, we elucidate the spatially resolved dynamics of electrons coupled to pumped phonons.  Our main results are: \\
 {{\bf \em 1. Phonon-induced disorder:}} We observe fast growth of local electronic correlations after the application of the pump.  A dramatic flattening in the momentum dependence of charge, spin and pairing correlations rapidly follows, pointing to loss of electronic spatial phase coherence.  We find that disorder emerges as a result of the nature of the initial light-created coherent phonon {\em superposition} state whose dynamics is effectively governed by a Hamiltonian that approximately conserves phonon occupations.  The presence of quasi-conserved phonon constants of motion implies that electronic observables self-average over the different disordered phonon configurations of the initial state and possess no off-diagonal coherence between different phonon sectors. This provides a realization of the disorder-free localization mechanism recently discussed in the context of lattice gauge theories~\cite{ASmith1,ASmith2,ASmith3,ASmith4,2DDisorderfree} in which disorder emerges dynamically due to a similar intricate interplay between the properties of the initial state and the symmetries of the Hamiltonian. To understand this behavior, we derive an effective model whose behavior captures the qualitative features of the exact dynamics on transient timescales.  Our effective theory provides a natural framework within which disorder and electron localization arise in the dynamics, and provides a perspective for the short-time dynamics complementing  analysis of the long-time behavior where loss of phonon coherence and preservation of the Poisson-distributed diagonal eigenvalues of the initial coherent state density matrix characterize a random disorder potential, responsible for the destruction of phase coherence of the  normal state electronic correlations~\cite{NonLinear4Kennes}. \\
{{\bf \em 2. Correlated electron-phonon steady state:}} We provide evidence that the system evolves to a steady state at long times characterized by sizeable correlations between electrons and phonons. The early-time dynamics that follow the pump already indicate rapid growth of local, negative correlations between the electron density $\hat{n}$ and the oscillator quadratic displacement $\hat{X}^2$ at a given site and anticorrelations of $\hat{X}$ at adjacent sites, which signals a tendency towards charge flow between neighboring sites, resulting in enhanced double occupancy.  This dynamical process quenches the Friedel oscillations \cite{AshAndMerm} of the electron density profile, and manifests as a space-time dependent feature in the  density-density correlation function that spreads spatially outwards along a ``light-cone'' defined by the Fermi velocity \cite{CalabreseCardy}. Behind the light cone, very rapidly  the density-density correlation function becomes basically structureless, suggesting that the asymptotic  state possesses a large degree of randomness. At long times, we find an overall increase in the magnitude of the expectation value of the electron-phonon interaction term, implying evolution towards a strongly correlated long-time electron-phonon state.\\
{\bf \em 3. Dynamically induced strong-coupling behavior}: We compare the dynamical electronic behavior in response to a pump in the quadratic-coupling model against that in the linear (Holstein) counterpart.
We observe larger double occupancy and greater large-amplitude response of momentum-resolved correlation peaks in the quadratic model, indicating that in this model, in contrast to the more widely studied Holstein model, the drive pushes the system into a strong-coupling regime. This substantive dynamical response of the non-linearly coupled system implies the existence of non-equilibrium pathways to coherent induction of electronic phases not accessible in equilibrium, and highlights the importance of the quadratic coupling in irradiated materials.
\bigskip

\begin{figure*}
\centering
\includegraphics[width=\columnwidth]{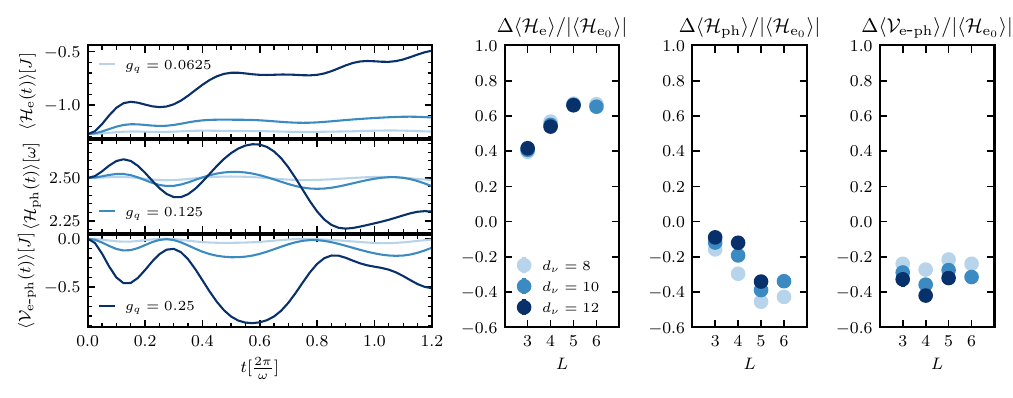}
\caption{{\bf Energy redistribution among the different system subsectors.} Infinite system iTEBD simulations (left panel) of the time dependence of the electronic (top), phononic (middle) and electron-phonon (bottom) energy densities for $\omega = \pi/2$ show a trend with larger $g_q$ of rapid heating of the electronic subsector, accompanied by transient relaxation of the electron-phonon subsector. Exact Krylov propagation of small systems  (right panel) with $L=3-6$ and local phonon Hilbert space dimension $d_{\nu} = 8, 10, 12$ ($L=6$ is restricted to $d_{\nu} =10$) for $\omega = \pi/2$ and the largest coupling strength $g_q=0.25$ to asymptotically long times showing the net change relative to the initial state in electronic (left), phononic (center) and electron-phonon (right) energy densities  confirms a correlated electron-phonon steady state, as evidenced by the considerable flow of energy from the electron-phonon subsector to the electronic subsector. The y-axis labels of the net change in energy densities have been placed at the top of the corresponding plots. Here $ {\mathcal H}_{{\rm e}_0}\equiv  {\mathcal H}_{{\rm e}} (0)$.}
 \label{fig:Energydensity}
\end{figure*}

\begin{figure*}
\centering
 \includegraphics[width=\columnwidth]{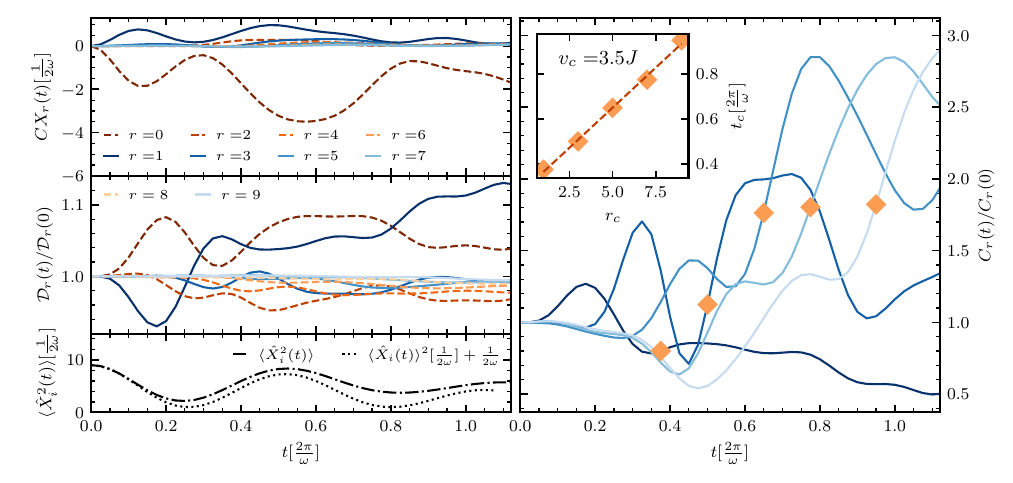} 
 \caption{{\bf Dynamics of charge and charge-phonon correlations.}  {\em Left column:} Time evolution of charge-lattice correlation $CX_r(t) = \langle \hat{n}_i \hat{X}_{i+r}^2 (t) \rangle - \langle \hat{n}_i (t) \rangle \langle \hat{X}_{i+r}^2 (t) \rangle$ (top) contrasted against that of $\langle \hat{X}_i^2(t) \rangle$ (bottom), and of the (connected) density-density correlation ${\mathcal{D}}_r(t) = \langle \hat{n}_i \hat{n}_{i+r} (t) \rangle$ normalized with respect to its initial-time value ${\mathcal{D}}_r(0)$ (middle).  Here $\hat{X}_i \coloneqq \sqrt{\frac{1}{2M \omega}} (b_i^\dagger + b_i)$, where $M$ is the oscillator mass, which we set to unity, $M=1$.  Note the violation of the relation $(\Delta {X}_i(t))^2 = \langle \hat{X}_i^2(t)\rangle - \langle \hat{X}_i (t)\rangle^2  = \frac{1}{2\omega}$ for $t \gtrapprox 0.15 \frac{2\pi}{\omega}$, an indication of deviation of the oscillator from an ideal coherent state.  {\em Right column:} Onset of a light-cone profile in the normalized density-density charge correlations; here $C_r (t)= \langle \hat{n}_i \hat{n}_j (t)\rangle -\langle \hat{n}_i(t) \rangle \langle \hat{n}_j (t)\rangle$ is normalized with respect to its initial-time metallic Friedel oscillations profile $C_r(0)$.   The diamond symbols mark the inflection point preceding the second maximum for the different $r$ lines, which we use in the inset to find a best fit of the light-cone charge propagation $t_c$ versus $r_c$ (dashed line), yielding an estimate for charge velocity:  $v_c \approx 3.5J$.  We use $g_q = 0.25$ and $\omega = \pi/2$ in this figure.
}  
 \label{fig:Pumpdynamics}
 \end{figure*}

\footnotetext[27]{This coupling also serves as a description of double-well electron-phonon systems \cite{BerciuNonLinear1,BerciuNonLinear2} and systems for which the linear approximation is inadequate \cite{Johnston}}

\noindent{\bf{\large{Formalism}}}\\
\noindent{\bf{Physical setup.}} We consider a metal whose vibrational modes are excited at initial time by a short-duration light pulse that creates a coherent phonon field \cite{Glauber} on every site, which couples non-linearly to the local electron density.  The Hamiltonian that governs the dynamics is given by
\begin{eqnarray} \label{Eq:QephHam}
 {\mathcal H} = {\mathcal H}_{\rm e} + {\mathcal{H}}_{\rm ph} + {\mathcal{V}}_{\rm e\mbox{-}ph}.
\end{eqnarray}
Here ${\mathcal H}_{\rm e} = - J \sum_{i,\sigma}  c_{i,\sigma}^\dagger c_{i+1,\sigma} + {\rm H.c.}$ characterizes the dynamics of electrons of spin flavor $\sigma \in \{\uparrow, \downarrow\}$ via the fermion creation (annihilation) operator $c_{i,\sigma}^\dagger$ ($c_{i,\sigma}$) and charge density $\hat{n}_i = \sum_\sigma \hat{n}_{i,\sigma}$ at site $i$.  The electrons of this irradiated system couple locally to the excited vibrations via the the dominant symmetry-allowed interaction \cite{NonLinear4Kennes,Note27}
\begin{eqnarray}  \label{eq:e-phCoupling}
{\mathcal V}_{\rm e\mbox{-}ph} = g_q \sum_i (\hat{n}_i - 1) (b_i^\dagger + b_i)^2.
\end{eqnarray}
The phonon Hamiltonian is $\mathcal{H}_{\rm ph} =  \omega \sum_{i} \left( b_i^\dagger b_i + \frac{1}{2} \right)$, which characterizes a local optical Einstein phonon mode with frequency $\omega~(\hbar = 1)$, described by the boson creation (annihilation) operator $b_i^\dagger$ ($b_i$). We set the lattice constant $a=1$ in what follows.

We simulate the time evolution of the initial state 
\begin{eqnarray}
\ket{{\mathbf \Psi}}= \ket{{\rm FS}}  \otimes \bigotimes_i \ket{\alpha}_i.
\end{eqnarray}
Here $\ket{\rm FS} = \prod_{k \leq k_{\rm F}} c^\dagger_{k,\uparrow} c^\dagger_{k,\downarrow} \ket{0}$ with $k_{\rm F} = \pi/2$ describes a metal formed from a Fermi sea of spinful electrons in a half-filled ($\langle \hat{n}_i \rangle = 1$) one-dimensional (1D) lattice and $\ket{\alpha} = {e^{-\frac{\abs{\alpha}^2}{2}}} \sum_{\nu} \frac{\alpha^{\nu}}{\sqrt{\nu !}} \ket{\nu}$
represents a coherent state of amplitude $\alpha$ written as an appropriate superposition of phonon-number states $\ket{\nu}$. Since the wavelength of the pump field extends beyond the lattice scale, we assume it produces a perfectly phase-coherent initial product state of onsite phonon coherent states $\bigotimes_i \ket{\alpha}_i$.

This model Hamiltonian implies an equilibrium renormalization of the oscillator stiffness $K \rightarrow K [1 + 4 \frac{g_q}{\omega} (\langle\hat{n}\rangle_i-1)]$. Thus, the onsite harmonic oscillator is stable so long as $\abs{g_q} <\frac{\omega}{4}$ \cite{NonLinear4Kennes} (see Supplementary Information). 
Here, we study the physics of the model for physical  parameters defined in units of $J$, i.e. we set $J=1$.
In the main text, we consider $g_q \leq 0.25$ for $\omega = \pi/2, \pi$ to study dynamics of the non-linear model for couplings ranging from weak to strong, and use $\alpha = \sqrt{2}$ for the pump amplitude.  This choice of $\omega$  allows us to numerically resolve the quantum effects in dynamics due to a large yet amenable phonon Hilbert space.
\medskip

\noindent{\bf{Simulations: Correlated electron-phonon steady state, electron density-density light-cone, and featureless electronic correlations.}} We simulate the time evolution of $\ket{\bf \Psi}$ in an infinite system on transient timescales via the infinite time-evolved block decimation (iTEBD) algorithm \cite{iTEBD} and access its long-time behavior in finite-size systems of size $L=3-6$ and local phonon Hilbert space dimension $d_{\nu} = 8, 10, 12$ using direct Krylov subspace methods. In iTEBD, one employs a  matrix-product state (MPS) ansatz for quantum states in the thermodynamic (infinite-size) limit, which permits access to information pertaining to long-ranged correlations in the system. Time evolution of an MPS is however ultimately limited to finite times because of the exponential growth of entanglement associated with a quench. Krylov subspace methods, based on Hamiltonian matrix-state vector multiplication, are in contrast not limited to short times, but are instead restricted to small $L$ due to the exponential growth of the Hilbert space with $L$.  Combining the two approaches allows use to derive reliable conclusions about long-range correlations on finite timescales from iTEBD and local correlations at long times from Krylov propagation.
\bigskip

\noindent{\bf{\large{Results}}}\\
Fig.~\ref{fig:Energydensity} demonstrates the energy redistribution amongst the different system subsectors in the course of the time evolution on timescales ranging from short (left panel) to long (right panel), as the system approaches its long-time limit of a correlated electron-phonon steady state. Consider the largest coupling $g_q = 0.25$ (dark lines in left panel). At early times $t \leq \frac{2\pi}{\omega}$, the electron subsystem absorbs energy from the excited phonons, and the phonon energy density oscillates about a value close to its initial value, while the electron-phonon energy density becomes more negative, see left panel of Fig.~\ref{fig:Energydensity}.  At asymptotically long times, we observe an overall flow of energy from the phonon and electron-phonon subsectors to the electron subsector (right panel of Fig.~\ref{fig:Energydensity}). Importantly, the increase in magnitude of the (negative) electron-phonon correlation term implies a  long-time correlated electron-phonon state.

Correlations between electrons and phonons already manifest in the early-time dynamics, as we demonstrate in Fig.~\ref{fig:Pumpdynamics}. Consider the charge-phonon correlation function $CX_r(t) = \langle \hat{n}_i \hat{X}_{i+r}^2 (t) \rangle - \langle \hat{n}_i (t) \rangle \langle \hat{X}_{i+r}^2 (t) \rangle$ (Fig.~\ref{fig:Pumpdynamics}, left; top), where $\hat{X}_i \coloneqq \sqrt{\frac{1}{2M \omega}} (b_i^\dagger + b_i)$, and we set $M=1$. For $r=0$, $\hat{n}$ rapidly becomes negatively correlated with $\hat{X}^2$. Note that $\langle \hat{n}_i (t) \rangle = 1$  throughout the dynamics in the translationally invariant system under consideration and $\langle \hat{X}_i^2(t)\rangle$ (dashdotted line, bottom) remains positive under time evolution. The substantial local, negative correlations in $CX_0(t)$ therefore imply a flow of electrons between neighboring sites. The same analysis applied to $CX_1(t)$ reveals a positive correlation between electron density and phonons separated by a single site with a dynamical profile somewhat similar (albeit of opposite sign) to $CX_0(t)$. With a slightly delayed onset, much weaker positive correlations build up at larger $r$ in $CX_r(t)$.  The interplay between onsite and nearest-neighbour correlations in $CX_r(t)$ reflects the tendency of charge to flow from a site to its neighbours, implying that doublons (doubly occupied sites) and holons (empty sites) emerge in the dynamics on such timescales. Indeed, in the middle panel, we observe a rapid enhancement of local electron density-density correlations  ${\mathcal{D}}_0(t) = \langle \hat{n}_i \hat{n}_i (t) \rangle = \langle \hat{n}_i \rangle + 2 \langle \hat{n}_{i,\uparrow} \hat{n}_{i,\downarrow}(t) \rangle$, accompanied by the suppression of ${\mathcal{D}}_1(t) = \langle \hat{n}_i \hat{n}_{i+1} (t) \rangle$ due to doublon creation, as expected if there is a tendency towards formation of an alternating pattern of doubly and singly occupied sites.  For times greater than $t \approx 0.175 [\frac{2\pi}{\omega}]$, ${\mathcal{D}}_1(t)$ begins to grow and becomes positive, whilst ${\mathcal{D}}_2(t) = \langle \hat{n}_i \hat{n}_{i+2} (t) \rangle$ diminishes, and a wavefront behavior in $r$ appears to arise.  In fact, when normalized against the $t=0$ metal Friedel density profile, a density-density correlation light-cone $C_r(t)/C_0(t)$ ($C_r (t)= \langle \hat{n}_i \hat{n}_j (t)\rangle -\langle \hat{n}_i(t) \rangle \langle \hat{n}_j (t)\rangle$)~\cite{CalabreseCardy,Kalthoff19} propagating outwards in $r$ can be clearly seen (Fig.~\ref{fig:Pumpdynamics}, right). A characteristic feature that emerges for larger $r$ at later time delays closely trails the second-in-time maximum. Thus, to sharply characterize the light-cone, we track the inflection point preceding that maximum (diamond symbols). A line of best fit through these data points (Fig.~\ref{fig:Pumpdynamics}, right; inset) reveals linear charge propagation with a velocity $v_c \approx 3.5 J$, slightly larger than the free metal Fermi velocity $2 k_{\rm F} J = \pi J$.  On the timescales accessed by iTEBD, we find no evidence for a  wave-front propagating in either of $CX_r(t)$ or $\langle \hat{X}_i (t) \hat{X}_{i+r} (t)\rangle$, reflecting the resistance to propagation of the  dispersionless Einstein oscillator modes of the initial-time ($g_q=0$) state. The behavior exhibited by $CX_r(t)$ and $C_r(t)$ implies non-equilibrium induction of enhanced double occupancy $\langle n_{i,\uparrow} n_{i,\downarrow}(t) \rangle$, which we have directly verified.

\begin{figure*}
\includegraphics[width=\columnwidth]{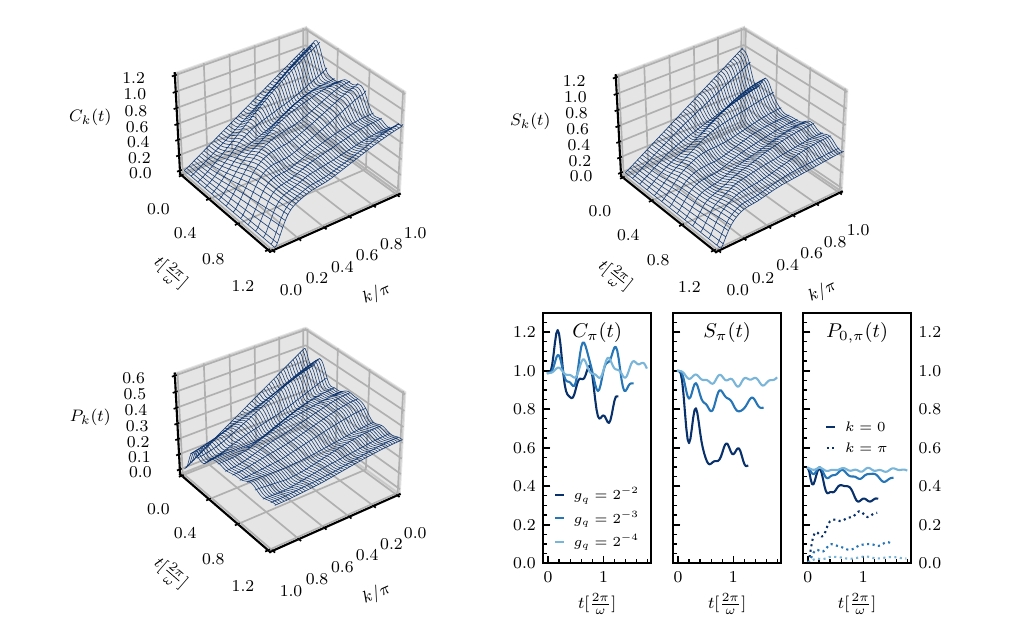}
\caption{{\bf Dynamics of momentum-resolved electronic correlations.} We study the evolution with time of momentum-resolved charge $C_k(t) = {\mathcal F}\{C_r(t)\}$, spin $S_k(t) = {\mathcal F}\{S_r(t)\}$ and pairing $P_k(t) = {\mathcal F}\{P_r(t)\}$ correlation functions for $g_q = 0.25$ and $\omega = \pi/2$ (three-dimensional plots) and the dependence on time of certain $k$ ($0,\pi$) correlations for various $g_q$ at $\omega = \pi/2$ (bottom, right). Note the $k$-axis of the $P_k(t)$ plot has been inverted for better visibility, and the y-axis labels of the $0$/$\pi$ correlations in the bottom right panel have been placed at the top of the corresponding plots.  Here $C_r \equiv \langle \hat{n}_{i}\hat{n}_{i+r} \rangle - \langle \hat{n}_{i} \rangle \langle \hat{n}_{i+r} \rangle, S_r \equiv \langle (\hat{n}_{i,\uparrow} - \hat{n}_{i,\downarrow})  (\hat{n}_{i+r,\uparrow} - \hat{n}_{i+r,\downarrow})\rangle$ and $P_r \equiv \langle c_{i,\uparrow}^\dagger c_{i,\downarrow}^\dagger c_{i+r,\downarrow} c_{i+r, \uparrow} \rangle$. ${\mathcal F}$ denotes the Fourier transform. Charge, spin and pairing correlations all rapidly flatten in the course of the dynamics. Note conservation of $C_0(t)$ and $S_0(t)$ in the dynamics.
}
 \label{fig:Flattening}
\end{figure*}

\begin{figure*}
\centering
 \includegraphics[width=\columnwidth]{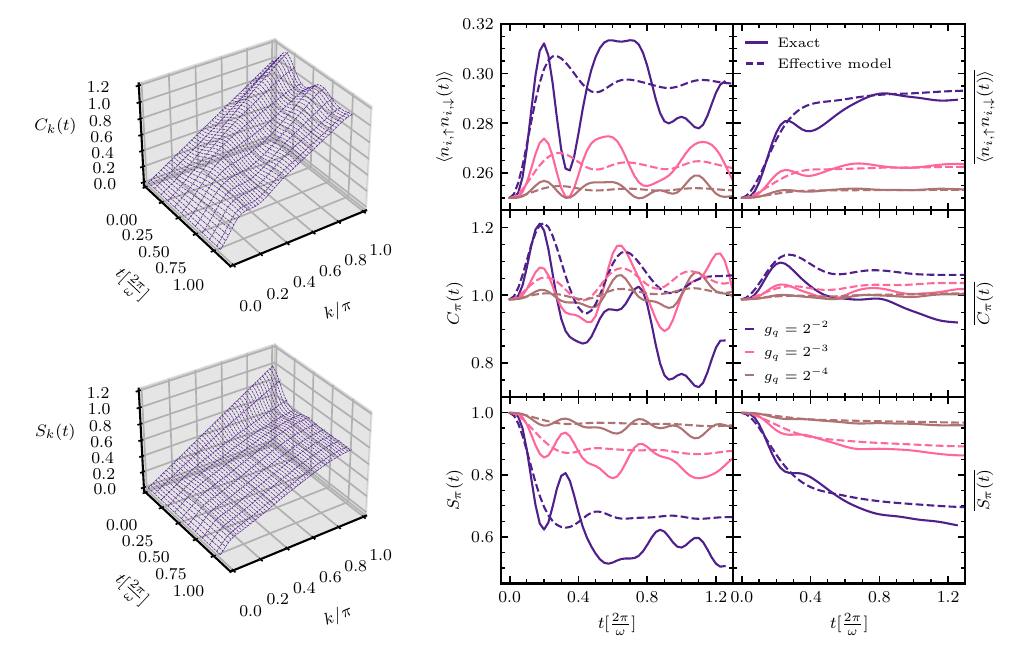}
 \caption{{\bf Dynamics of a pumped metal in the effective theory.} {\em Left column:}  Evolution with time of momentum-resolved charge $C_k(t) = {\mathcal F}\{C_r(t)\}$ ($C_r \equiv \langle \hat{n}_{i}\hat{n}_{i+r} \rangle - \langle \hat{n}_{i} \rangle \langle \hat{n}_{i+r} \rangle$) and  spin $S_k(t) = {\mathcal F}\{S_r(t)\}$ ($S_r \equiv \langle (\hat{n}_{i,\uparrow} - \hat{n}_{i,\downarrow})  (\hat{n}_{i+r,\uparrow} - \hat{n}_{i+r,\downarrow})\rangle$) correlation functions (${\mathcal F}$ denotes the Fourier transform) for $g_q = 0.25$ and $\omega = \pi/2$ (three-dimensional plots) in the effective model given by Eqs.~\eqref{Eq:Heff.}, \eqref{Eq:LinearResponse} from iTEBD simulations. {\em Right column:} Dependence on time of raw (left panels) and time-averaged (right panels) double occupancy $\langle \hat{n}_{i,\uparrow} \hat{n}_{i,\downarrow} (t) \rangle$, $\pi$-charge $C_{\pi}(t)$ and $\pi$-spin $S_{\pi}(t)$ correlations for various $g_q$ at $\omega = \pi/2$ in the exact model (solid line) and the effective  model given by Eqs.~\eqref{Eq:Heff.}, \eqref{Eq:LinearResponse} (dashed line) from iTEBD simulations. A bar label over an observable symbol denotes time averaging: $\overline{\langle \hat{O}(t) \rangle} = \frac{1}{t} \int_0^t {\rm d} \tau \langle \hat{O} (\tau) \rangle $. We observe good agreement between results obtained in the effective model and the exact simulations of the fully coupled model, including the rapid flattening of charge and spin correlations in the course of the dynamics.
 }
 \label{fig:Disorder}
 \end{figure*}

Turning to the dynamics of long-range electronic correlations, in Fig.~\ref{fig:Flattening} we study the evolution with time of the momentum-resolved charge $C_k(t)$, spin $S_k(t)$ and pairing $P_k(t)$ correlation functions to fully characterize the electronic features.
 Apart from a fast initial growth of $C_{\pi}(t)$ for $t \lessapprox 0.2 [\frac{2\pi}{\omega}]$ due to the enhanced double occupancy, we observe rapid flattening in momentum space of these correlations, marking the loss of spatial  coherence, despite the persistent growth of local density-density and charge-phonon correlations, indicating that the pattern of doubly and singly occupied sites is becoming random. This surprising behavior implies an effective {\em disordered} state forms on transient timescales, and a more subtle role played by phonons in the dynamics, as we show next.
\medskip

\noindent{\bf{Effective model for disorder.}} To understand the mechanism behind the appearance of disorder, we derive an effective theoretical picture for the dynamics  to leading order in $g_q/\omega$.  

For reasons that will become apparent in what follows, we find it convenient to consider a rotating frame in which the off-diagonal phonon terms  (in the occupation-number  basis) of Eq.~\eqref{eq:e-phCoupling} are eliminated via a Bogoliubov-type squeezing transformation
\cite{NonLinear4Kennes}:  ${\mathcal H}  \rightarrow {\tilde{\mathcal H}} =  U {\mathcal H} U^{\dagger}$, where $U = e^S$, $S=-\sum_j \frac{1}{2}\zeta_j (b^{\dagger}_j b^{\dagger}_j  - b_j b_j)$, and $\zeta_i = -\frac{1}{4} \ln[1+4\frac{g_q}{\omega}(\hat{n}_i-1)]$, the squeezing parameter, is chosen so that the $(b^{\dagger}_i)^2$ and $(b_i)^2$ terms vanish.  This yields, in the squeezed frame, $
 {\tilde{\mathcal H}}  = e^S {\mathcal H}_{\rm e} e^{-S} + \sum_i \omega \sqrt{1+4\frac{g_q}{\omega}(\hat{n}_i-1)} (\beta^\dagger_i \beta_i + \frac{1}{2})$, where $\beta^\dagger$ creates a squeezed phonon state. Perturbatively expanding the transformed coupling term in orders of $g_q/\omega$, we find 
\begin{eqnarray} \label{Eq:Heff.}
  {\mathcal H_{\rm eff.}} &= -  &J^*  \sum_{i,\sigma}  (c_{i,\sigma}^\dagger c_{i+1,\sigma} + {\rm H.c.}) + \omega^* \sum_i  \left(\beta^\dagger_i \beta_i + \frac{1}{2}\right)  \nonumber \\
  && + 2 g_q  \sum_i  (\hat{n}_i-1) \left(\beta^\dagger_i \beta_i + \frac{1}{2} \right)  \nonumber \\
  && - 4 \frac{g_q^2}{\omega} \sum_i  (\hat{n}_{i,\uparrow} - 1/2) (\hat{n}_{i,\downarrow} - 1/2) \left(\beta^\dagger_i \beta_i + \frac{1}{2} \right). 
\end{eqnarray}
Here $J^* = J e^{-\frac{1}{2}(\frac{g_q}{\omega})^2 (\langle \hat{n}_\mathrm{B}\rangle ^2 + 2\langle \hat{n}_\mathrm{B} \rangle + 1)}$ ($\langle \hat{n}_\mathrm{B} \rangle$ is the average number of excited bosons in the dynamics) and $\omega^* = \omega - g_q^2/\omega$. Aside from renormalization of the electron and phonon energy scales, we see that the electron density, at $\mathcal{O}\{g_q/\omega\}$, and double occupancy, at $\mathcal{O}\{(g_q/\omega)^2\}$, couple to the squeezed phonon density.  This Hamiltonian is exact to  $\mathcal{O}\{g_q/\omega\}$, and approximate to  $\mathcal{O}\{(g_q/\omega)^2\}$ (and higher orders). See Supplementary Information for details of the derivation and approximations employed. For the time dependence of electronic operators $\hat{O}_{\rm e}$ measured in the original frame, we derive in a similar approximation (details in the Supplementary Information) a theory in the squeezed frame in which $\hat{O}_{\rm e}$ transforms as $\hat{O}_{\rm e} \rightarrow e^S\hat{O}_{\rm e} e^{-S}$, the initial state as $ \ket{0} \equiv \ket{{\mathbf \Psi}} \rightarrow e^S\ket{0}$, and ${\mathcal U}_{\rm eff.} = e^{-i{\mathcal H_{\rm eff.}}t}$ governs the time evolution. Within this scheme in which  terms larger than $\mathcal{O}\{g_q/\omega\}$ are neglected, the equal-time expectation value of  $\hat{O}_{\rm e}$ in the squeezed frame becomes
\begin{eqnarray} \label{Eq:LinearResponse}
\langle \hat{O}_{\mathrm{e}} (t)\rangle  &= \bra{0} {\mathcal U}^\dagger_{\rm eff.}(t) \hat{O}_{\mathrm{e}}  &{\mathcal U}_{\rm eff.}(t) \ket{0} \nonumber \\
&&+ \bra{0} {\mathcal U}^\dagger_{\rm eff.} (t) \Gamma_{\hat{O}_{\mathrm{e}}}  {\mathcal U}_{\rm eff.}(t)  \ket{0},
\end{eqnarray}
with $\Gamma_{\hat{O}_{\mathrm{e}}} = \comm{S}{\hat{O}_{\mathrm{e}}}$. In Fig.~\ref{fig:Disorder}, we test the predictions of Eq.~\eqref{Eq:LinearResponse} against the exact results. Not only does the effective theory reproduce the flattening in momentum-resolved $C_k(t)$ and $S_k(t)$ (Fig.~\ref{fig:Disorder}, left) observed in the exact simulations (Fig.~\ref{fig:Flattening}), it also provides an overall qualitative proxy for the exact raw and time-averaged quantities $\langle n_{i,\uparrow} n_{i,\downarrow}(t) \rangle$, $C_{\pi}(t)$ and $S_{\pi}(t)$ (Fig.~\ref{fig:Disorder}, right), even for relatively large $g_q/\omega$ for which the approximations we employ are less justified (we discuss limitations of the effective model in the Supplementary Information).

The origin of disorder becomes manifest in the effective model.  Noting that in Eq.~\eqref{Eq:LinearResponse} $\Gamma_{C_k}, \Gamma_{S_k} = 0$ because charge and spin correlations are conserved under the squeezing transformation implies that the exact dynamics is approximately captured by an effective theory that conserves the squeezed phonon occupations. This effective theory thus encodes dynamics of the electrons within independent trajectories of different squeezed phonon configurations in an ensemble given by an initial Poisson-distributed linear combination that describes the $t=0$ state (now in the squeezed basis) and is thus formally equivalent to the disorder-averaged dynamics of an electronic system quenched in a random, static Poisson-distributed potential determined by the initial state occupations. Charge and spin correlations, by construction, possess no coherence between different squeezed phonon sectors, and thus very quickly flatten  in the course of the dynamics.  The exact electronic behavior on transient timescales is therefore dominated by a large degree of effective disorder despite that the initial state and the Hamiltonian in both squeezed and unsqueezed frames are disorder free.

Note however while this effective model remains valid on intermediate timescales, higher-order terms in $g_q/\omega$, neglected in our treatment, eventually become important, which may lead to deviations from the above behavior. Nonetheless, our numerics seem to suggest evolution towards a state with large disorder that remains robust for extended timescales.  We provide  in the Supplementary Information a complementary treatment of electronic disorder at later times in the unsqueezed frame based on the dynamics induced by phonon decoherence.  This disordered behavior persists despite the attractive electron density-density interaction term of  ${\mathcal H_{\rm eff.}}$, which, at least in 1D, implies that the system lies within a regime far from the superfluid transition~\cite{Fisher1DSI}.

The picture we obtain here indicates that a translationally uniform system excited by a spatially uniform field governed by electron-phonon non-linearity will flow towards a state characterized by a high level of randomness in absence of quenched disorder.  This behavior was noted in Ref.~\cite{NonLinear4Kennes} based on an analysis of phonon decoherence (see Supplementary Information) and has become a subject of major theoretical interest within the field of disorder-free localization~\cite{Grover,ASmith1}. In this regard, our effective theory reveals a mechanism operative in experiment for dynamically induced disorder reminiscent of that found in the context of special models of lattice gauge theory~\cite{ASmith1,ASmith2,ASmith3,ASmith4,2DDisorderfree}. These models describe the coupling of fermions to background gauge fields modeled as spin degrees of freedom in which a duality transformation~\cite{Duality1,Duality2} maps the Hamiltonian onto one with conserved  gauge charge configurations and  the gauge charge couples directly onsite to the fermion occupation.  Time evolution with this manifestly translationally invariant Hamiltonian of an initial product state of fermions and gauge spins, equivalent to a linear superposition over different superselection gauge charge configurations, exhibits disorder-free localization due to self-averaging of observables over the different initial gauge configurations. In contrast to these models, our theory reveals that an {\em approximate} effective model governed by similar behavior dominates the exact dynamics of the quenched electron-phonon system  on extended timescales. Thus, our work paves a way towards physical realization of this phenomenon in the realistic setup of current pump-probe experiments. Furthermore, the emergence of an attractive Hubbard interaction in the effective model presents an unexplored avenue within the context of disorder-free localization to study the competition between disorder and attractive interactions in the dynamics of spinful fermionic systems. 

\medskip

\begin{figure}
\centering
\includegraphics[width=\columnwidth]{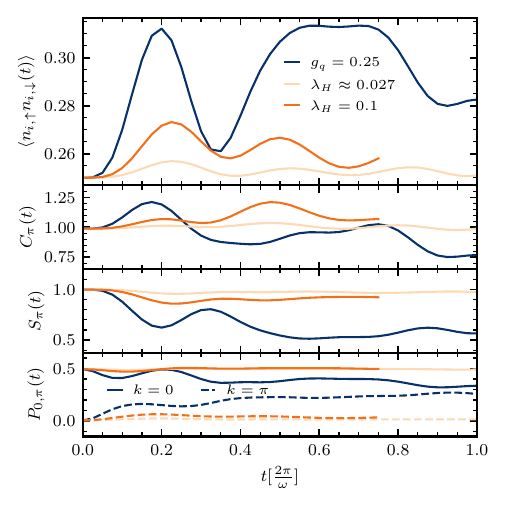}
\caption{{\bf Dynamical response in the quadratically coupled model versus in the Holstein model.}  A comparison of the pump-induced dynamics in the quadratic-coupling model with coupling constant $g_q$ to that of the Holstein model with coupling constant $g_H$ and dimensionless effective coupling parameter $\lambda_H = \frac{g_H^2}{2 \omega J}$ for appropriately selected values of the couplings and $\omega = \pi/2$ (see text and Supplementary Information for more details) reveals that the driven quadratic model induces a more appreciably enhanced double occupancy (uppermost panel) and causes a greater response in electron correlations (lower three panels) including the flattening of pairing tendencies (lowermost panel) than its Holstein model counterpart.}
 \label{fig:QvH}
\end{figure}

\noindent{\bf{Comparison with a linearly coupled electron-phonon model: Dynamically induced strong-coupling behavior from non-linear electron-phonon coupling.}} Before we conclude, we contrast the dynamics of our non-linear model to that of the (linear) Holstein model (which cannot be induced by a light pulse in an inversion symmetric system). We use two methods to choose an appropriate  coupling strength in the Holstein model corresponding to a given coupling strength of the quadratic model against which we  perform a comparison, see Supplementary Information for details. In one approach we choose the Holstein coupling that yields the same equilibrium ground state double occupancy as in the quadratic model. In the other the Holstein coupling is chosen to produce the same double occupancy as that obtained analytically from a disentangling transformation  that serves as a low-energy description of the dynamics (Eq.~\eqref{Eq:Heff.}). Both methods of comparison show that even a relatively large non-linear coupling such as $g_q = 0.25$, proximate to the oscillator instability threshold, gives rise in equilibrium to weak-coupling behavior. In contrast, Fig.~\ref{fig:QvH} shows that the quadratic model exhibits a much stronger dynamical response to the pump, displaying both a large enhancement of double occupancy (uppermost panel), and  large-amplitude dynamics in momentum-resolved electronic correlations (lower panels) including flattening of $P_k(t)$  (lowermost panel).  This is in sharp contrast to the behavior of the Holstein model and implies that dynamics of non-linear coupled electron-phonon systems operative in pump-probe experiments affords non-equilibrium pathways to correlated physics unavailable in the static limit and which lies outside the frame of conventional theoretical models.
\bigskip

\noindent{\bf{\large{Discussion}}}\\
Prior studies of non-linear electron-phonon dynamics have relied on approximate low-energy treatments.  Our exact numerical approach to spatially resolved dynamics of a pumped non-linear electron-phonon systems fills an urgent need. We use iTEBD to provide a detailed exact analysis of short-time (up to $t\sim \frac{2\pi}{\omega}$) dynamics of an infinite non-linear electron-phonon coupled metal upon coherent excitation of vibrational modes by light.  We supplement this by direct Krylov propagation of small systems to asymptotically long times.  We explicitly describe the flow towards a correlated electron-phonon steady state at long times, the indication of which already manifests on short timescales.  Remarkably, although we consider a spatially uniform system evolving after application of a  spatially uniform pump field, the key feature of the long-time state is the appearance of properties consistent with a high degree of effective disorder that dominates the physical behavior, unveiling an intriguing connection to the scenario of disorder-free localization~\cite{ASmith1,ASmith2}. These properties are a consequence of the quasi-conserved squeezed-phonon constants of motion that effectively govern the time evolution of the initial linear superposition state and the very rapid loss of coherence of the phonons, which we found to be directly tied to the buildup of disorder, implying that the intermediate- and long-time state is an incoherent superposition of different oscillator configurations on different sites.   These incoherent phonon configurations result in a dynamic effective disorder potential for the electrons, which leads to the suppression of the (power-law) quasi-long-range charge, spin and pairing correlations. Analysis of the energy redistribution amongst the different system subsectors and of electron and phonon distribution functions of the long-time state obtained in finite-size systems, presented in the Supplementary Information, suggests that the terminal state obtained in finite-size simulations may not be thermal (in the Eigenstate thermalization hypothesis (ETH)~\cite{ETH} sense). Determining the fate of the established long-time entangled electron-phonon state in which the phonons in effect provide strong onsite potential fluctuations that substantially broaden all momentum-space distribution functions and fully disentangling the contributions of electron heating from  localization due to the transient phonon-induced disorder to this entangled electron-phonon state are beyond the scope of this paper, and are left to future work.

A crucial question, not resolved by this work, relates to the possibility of pump-induced superconductivity as predicted in Ref. \cite{NonLinear4Kennes}. In our calculations no evidence for superconductivity is found and we only find weak evidence for charge density wave correlations for very short time delays; the results are more consistent with the system falling within the disorder-dominated Anderson insulating regime of the phase diagram presented in Ref. \cite{NonLinear4Kennes}. One possibility would be that superconducting and density wave regimes either do not exist or are not accessible with the current pump protocol (perhaps because the pump transfers too much energy to the electronic subsystem). A second possibility would be that the one-dimensional model considered here disfavors superconductivity.  In fact, it has been shown that quantum fluctuations can destroy superconductivity in dirty superconductors below a mobility threshold \cite{MAHLEE}. In one dimension, all single-particle states are localized in presence of a static disorder potential.  Despite that in one-dimensional systems superconductivity can overcome the localizing tendency of disorder to some extent \cite{GiammarchiSchulz}, the effects of disorder are stronger than in higher dimensions.   The accurate simulation of pump-induced dynamics in higher-dimensional systems in the thermodynamic limit faces challenges, but is urgently needed.

The  quadratic model reacts more strongly to a pump than the linear Holstein model, highlighting the importance of this mechanism in pump-probe experiments, e.g. \cite{Lanzara}.  These  results thus generally apply to light irradiated centrosymmetric crystals. Questions such as the consideration of additional electron-vibration interactions consistent with inversion symmetry \cite{EcksteinNonLinear,LightBP}, which may aid in the stabilization of a transient superconducting state, as well as how the the electron-phonon steady state exposed in this work manifests experimentally are also important open challenges and call for the development of new tools for the study of out-of-equilibrium non-linear electron-phonon problems. An intriguing possibility is to use the information obtained here about the properties of the long-time state to motivate a variational ansatz in order to simulate the dynamics.

The pump-activated {\em transient} phonon-induced disorder in electron dynamics presents an opportunity to explore the interplay between correlations and randomness in out-of-equilibrium electronic matter.
\bigskip

\noindent{\bf{\large{Methods}}}\\
We study pump-induced dynamics via exact numerical simulations of the non-linear model coupled with an effective theory derived within a treatment formally similar to a linear response theory in a low-order expansion in powers of $g_q/\omega$.\\

\noindent{\bf Details of exact numerical simulations of the non-linear electron-phonon system.} We simulate the time evolution of $\ket{\bf \Psi}$ representing the metal on an infinite chain irradiated at initial time $t=0$ by a pump via the infinite time-evolved block decimation (iTEBD) algorithm \cite{iTEBD} utilizing the TeNPy Library \cite{TeNPy}.  We use $d_{\nu} = 12$ phonon states to represent the local phonon Hilbert space. We allow the bond dimension $\chi$ to grow without saturation in the iTEBD time evolution, and converge our results with respect to the truncation error $\epsilon_{\rm TEBD}$.  This allows access to time $t \sim 5J$ for which we find $\epsilon_{\rm TEBD} = 10^{-3.5}$ achieves satisfactory convergence.  We refer the reader to Supplementary Information for more information. To  shed light on the long-time behavior we also propagate the initial state using direct Krylov subspace methods for finite system sizes $L=3-6$ with $d_{\nu} = 8, 10, 12$ and twisted boundary conditions, see Supplementary Information for more details.\\

\noindent{\bf Details of effective model obtained within a low-order expansion in $g_q/\omega$.}
We derive an effective model within a framework similar to linear response, consistently incorporating contributions of $\mathcal{O}\{g_q/\omega\}$, with judiciously selected $\mathcal{O}\{(g_q/\omega)^2\}$ corrections (e.g., the effective electron density-density interaction term).  This theory, strictly valid to $\mathcal{O}\{g_q/\omega\}$, qualitatively captures the exact behavior of the time-evolved initial state in infinite systems obtained using iTEBD.  We simulate the dynamics governed by the effective model by time-evolving $\ket{0}\equiv \ket{\bf \Psi}$ under the action of ${\mathcal U}_{\rm eff.}(t)$ in Eq.~\eqref{Eq:LinearResponse} using iTEBD, employing $d_{\nu} = 12$ local squeezed phonon states and allowing $\chi$ to grow without saturation in the time evolution, while converging results with respect to $\epsilon_{\rm TEBD}$.  This allows access to time $t \sim 5J$ for which we find $\epsilon_{\rm TEBD} = 10^{-3.5}$ achieves satisfactory convergence. Details of the derivation of the effective model and additional discussion of the dynamics are presented in the Supplementary Information.

\noindent{\bf{\large{Acknowledgments}}}\\
\noindent J.~S., D.~R.~R. and A.~J.~M. acknowledge support from the National Science Foundation (NSF) Materials Research Science and Engineering Centers (MRSEC) program through Columbia University in the Center for Precision Assembly of Superstratic and Superatomic Solids under Grant No. DMR-1420634. B.~K. acknowledges support from NSF Grant No. CHE-1954791. D.~M.~K. acknowledges support from Deutsche Forschungsgemeinschaft (DFG, German Research Foundation) under Germany's Excellence Strategy - Cluster of Excellence Matter and Light for Quantum Computing (ML4Q) EXC 2004/1 - 390534769 and from the Max Planck-New York City Center for Non-Equilibrium Quantum Phenomena. This work used the Extreme Science and Engineering Discovery Environment (XSEDE), which is supported by NSF grant number ACI-1548562, through allocation TG-DMR190074. J.~S. also acknowledges the hospitality of the Center for Computational Quantum Physics (CCQ) at the Flatiron Institute. The Flatiron Institute is a divison of the Simons Foundation.



\bibliography{article}




\clearpage
\break

\onecolumngrid

\pagenumbering{arabic}

\setcounter{figure}{0} 
\captionsetup[figure]{name=Fig.}
\renewcommand\thefigure{S\arabic{figure}}

\section*{\bf{\large{Supplementary Information}}}

\renewcommand{\thesection}{\Roman{section}}

\section{Stability of the quadratic electron-phonon model}\label{Stabapp}

In this section, we derive the stability condition of the quadratically coupled electron-phonon model we consider. The Hamiltonian of the model reads
\begin{eqnarray}
{\mathcal H} = {\mathcal H}_{\rm e} + {\mathcal{H}}_{\rm ph} + {\mathcal{V}}_{\rm e\mbox{-}ph},
 \label{EqS1}
\end{eqnarray}
where
\begin{eqnarray} 
&& {\mathcal H}_{\rm e} = - J \sum_{i,\sigma}  c_{i,\sigma}^\dagger c_{i+1,\sigma} + {\rm H.c.},  \label{EqS1details1}\\
&& \mathcal{H}_{\rm ph} =  \omega \sum_{i} \left( b_i^\dagger b_i + \frac{1}{2} \right), \label{EqS1details2} \\
&& {\mathcal V}_{\rm e\mbox{-}ph} = g_q \sum_i (\hat{n}_i - 1) (b_i^\dagger + b_i)^2.
 \label{EqS1details3}
\end{eqnarray}

A stable harmonic oscillator mode localized on a given site implies an oscillator stiffness $K>0$.  To derive the condition for stability of the coupled electron-phonon system, we rewrite  ${\mathcal H}$ in terms of the harmonic oscillator displacement $\hat{X}_i$ and momentum $\hat{P}_i$ operators.  We make use of the relation ($\hbar = 1$)
\begin{eqnarray}
b_i = \sqrt{\frac{M \omega}{2}} (\hat{X}_i + \mathrm{i} \frac{1}{M \omega} \hat{P}_i),
\end{eqnarray}
where $M$ is the oscillator mass, to obtain
\begin{eqnarray}
&&\mathcal{H}_{\rm ph} = \sum_i \frac{1}{2} K \hat{X}_i^2 + \sum_i \frac{1}{2M} \hat{P}_i^2  \\
&& {\mathcal V}_{\rm e\mbox{-}ph} = 2  \frac{g_q}{\omega} K \sum_i (\hat{n}_i - 1) \hat{X}_i^2,
\end{eqnarray}
in which we used $K = \omega^2 M$.  Thus, the quadratic electron-phonon coupling renormalizes the oscillator stiffness on any given site 
\begin{eqnarray}
K \rightarrow K \Big[1+4\big( \hat{n}  - 1\big)\frac{g_q}{\omega}\Big].
\end{eqnarray}
Demanding that $K>0$, we arrive at the stability condition of the electron-phonon model:
\begin{eqnarray}
\abs{g_q}< \frac{\omega}{4}.
\end{eqnarray}
For spinless electrons $\hat{n}-1 \rightarrow \hat{n}-1/2$ and the stability condition, then, is $\abs{g_q}< \frac{\omega}{2}$.

\section{Effective model}
In this section, we derive in detail the effective model for the disordered dynamics we obtain within a treatment, reminiscent of  linear response, based on a low-order expansion in $g_q/\omega$ (and valid for arbitrary quench amplitude).  We discuss our approximations and limitations of the model.

As discussed in the main text, we move to a rotating squeezed frame, in which we derive a response theory to leading order in $g_q/\omega$. By direct comparison against the exact results we show that the effective model captures the main qualitative features of the exact model dynamics.

\subsection{Squeezing transformation}\label{Sqtr}
Kennes et al. \cite{NonLinear4Kennes} found a transformation that rescales the phonon coordinate, rotating the Hamiltonian Eqs.~\eqref{EqS1}-\eqref{EqS1details3} into a frame in which the electrons and phonons are approximately decoupled. The electron density-dependent transformation  ${\mathcal H}  \rightarrow {\tilde{\mathcal H}} = U {\mathcal H} U^{\dagger}$, with $U = e^S$, $S=-\sum_j \frac{1}{2}\zeta_j (b^{\dagger}_j b^{\dagger}_j  - b_j b_j)$ and squeezing parameter $\zeta_i = -\frac{1}{4} \ln[1+4\frac{g_q}{\omega}(\hat{n}_i-1)]$, yields 
  \begin{eqnarray}
  &&\beta^\dagger_i \equiv e^S b^\dagger_i e^{-S} = \cosh(\zeta_i) b^\dagger_i + \sinh(\zeta_i) b_i, \nonumber \\
  &&\beta_i \equiv e^S b_i e^{-S}= \cosh(\zeta_i) b_i + \sinh(\zeta_i) b^\dagger_i.
  \end{eqnarray}
  Here $\beta^\dagger_i$ creates a squeezed phonon state on site $i$. Under this transformation, ${\mathcal{H}}_{\rm ph} + {\mathcal{V}}_{\rm e\mbox{-}ph}$ is recast into a form completely diagonal in the squeezed phonon occupation basis:
\begin{eqnarray}
{\tilde{\mathcal H}} = {\tilde{\mathcal H}}_{\rm e} + \sum_i \omega \sqrt{1+4\frac{g_q}{\omega}(\hat{n}_i-1)} (\beta^\dagger_i \beta_i + \frac{1}{2}),
\end{eqnarray}
where ${\tilde{\mathcal H}}_{\rm e} = e^S {\mathcal H}_{\rm e} e^{-S} $ is the squeezed electronic Hamiltonian (we discuss a treatment of this term below).  Our formulation of the problem  amounts to a quench in which the non-linear electron-phonon coupling is suddenly switched on. (We have verified that the quench dynamics of electronic correlations obtained from the time evolution of the initial product state of electrons and phonons exactly resembles that obtained from an initial correlated electron-phonon state on the accessible timescales in iTEBD.)  In the original frame this generates phonon quanta in the dynamics due to the off-diagonal phonon terms in the coupling (Eq.~\eqref{EqS1details3}). In contrast, the squeezed Hamiltonian conserves the boson number, which, however, couples to the square-root of the electron density operator.  We will see next that we can take advantage of this form in order to understand the nature of the dynamics of electronic correlations.  

The ratio of energy scales $g_q/\omega$ arises naturally in the squeezed frame. This provides an opportunity to expand the interaction term in the squeezed Hamiltonian directly in orders of $g_q/\omega$. In the limit $g_q\ll\omega$, we Taylor expand to $\mathcal{O}\{(g_q/\omega)^2\}$ and find
\begin{eqnarray}
{ \tilde{\mathcal H}} &\approx& {\tilde{\mathcal H}}_{\rm e} + \sum_i \big[\omega - 2 \big(g_q + \frac{g_q^2}{\omega} \big) \big] (\beta^\dagger_i \beta_i + \frac{1}{2})  +  2 \big(g_q + \frac{g_q^2}{\omega} \big)  \sum_i   \hat{n}_i(\beta^\dagger_i \beta_i + \frac{1}{2}) -  4 \frac{g_q^2}{\omega} \sum_i \hat{n}_{i,\uparrow} \hat{n}_{i,\downarrow} (\beta^\dagger_i \beta_i + \frac{1}{2}).
\label{EqSbeta}
\end{eqnarray}
To this order the electron-phonon coupling is completely local and the squeezed phonon number on each site is conserved.  The second term describes a squeezed phonon bath term with a renormalized energy scale $\tilde{\omega} = \big[\omega - 2 \big(g_q + \frac{g_q^2}{\omega} \big)\big]$, which includes corrections at both $\mathcal{O}\{g_q/\omega\}$ and $\mathcal{O}\{(g_q/\omega)^2\}$. The third term shows that the phonon occupation on site $i$ changes the electron's local chemical potential, again at both orders.  This gives rise to disorder, static at this level of approximation, due to the nature of the initial state (as explained within our effective theory, see below and main text). Higher-order terms neglected in the transformation will  lead to the evolution of $\beta_i^\dagger \beta_i$, changing the disorder from static to dynamic. Phonons also mediate an effective local electron-electron attraction (fourth term) that appears first at second order.

We have so far postponed a treatment of the ${\tilde{\mathcal H}}_{\rm e}$. To proceed, we evaluate this term in the electron occupation number basis:
\begin{eqnarray}
{\tilde{\mathcal H}}_{\rm e} \equiv {\tilde{\mathcal H}}_{\rm e} [\{ n_i, n_j \}] = -J \sum_{\langle ij \rangle, \sigma} c_{i,\sigma}^\dagger c_{j,\sigma} e^{ (\zeta[n_i + 1] - \zeta[n_i] )\mathcal{B}_i} e^{(\zeta[n_j - 1] - \zeta[n_j] )\mathcal{B}_j},
\label{EqExponentialsInIncoherent}
\end{eqnarray}
where $\langle ij \rangle$ refers to $j = i \pm 1$ and $\mathcal{B}_i \equiv \frac{1}{2} (\beta_i^\dagger \beta_i^\dagger - \beta_i \beta_i)$.  Expanding the logarithm in the definition of $\zeta_i$ to $\mathcal{O}\{g_q/\omega\}$, we find $\zeta[n_i \pm 1] - \zeta[n_i] = \mp g_q/\omega$. Thus, to $\mathcal{O}\{g_q/\omega\}$, the exponential factors in Eq.~\eqref{EqExponentialsInIncoherent} reduce to unity and we retrieve the original untransformed electronic hopping term.  To incorporate corrections in the electronic hopping due to the squeezing transformation we must expand the exponentials to next order:
\begin{eqnarray}
e^{(\zeta[n_i + 1] - \zeta[n_i] )\mathcal{B}_i} e^{ (\zeta[n_j - 1] - \zeta[n_j] )\mathcal{B}_j} \approx 1 + (\zeta[n_i + 1] - \zeta[n_i] ) && (\zeta[n_j - 1] - \zeta[n_j] ) \mathcal{B}_i \mathcal{B}_j  \nonumber \\
&&+ \frac{1}{2} \big( (\zeta[n_i + 1] - \zeta[n_i] )^2 \mathcal{B}_i^2 +  (\zeta[n_j - 1] - \zeta[n_j] ) \mathcal{B}_j^2 ) \big).\nonumber
\end{eqnarray}
Invoking an inelastic approximation in which one neglects correlations between phonon states on different sites and thus the $\mathcal{B}_i \mathcal{B}_j$ term (this can be rationalized either by simply noting that the initial phonon state is a product over site wavefunctions whose inter-site correlations ought to be unimportant at very early times or by utilizing an incoherent approximation to the phonon density matrix~\cite{NonLinear4Kennes}), and evaluating the last term in the initial coherent phonon state, we find a contribution at $\mathcal{O}\{(g_q/\omega)^2\}$: $-\frac{1}{2}(\frac{g_q}{\omega})^2 (\langle \hat{n}_{\rm B} \rangle^2 + 2 \langle \hat{n}_{\rm B} \rangle + 1)$, where $\langle \hat{n}_{\rm B} \rangle = \abs{\alpha}^2$ is the expectation value of $\hat{n}_{\rm B} = \beta^\dagger \beta$ in the initial  state. Re-summing all similar contributions in the exponential, we  find 
$$J^* = J e^{-\frac{1}{2}(\frac{g_q}{\omega})^2 (\langle \hat{n}_\mathrm{B}\rangle ^2 + 2\langle \hat{n}_\mathrm{B} \rangle + 1)}.$$
This result is exact to $\mathcal{O}\{g_q/\omega\}$ and approximate to  $\mathcal{O}\{(g_q/\omega)^2\}$ and higher orders due to the inelastic approximation. 

Collecting the various terms, we arrive at the approximate effective Hamiltonian, Eq.~\eqref{Eq:Heff.} of the main text:
 \begin{eqnarray} \label{Eq:Heff.SI}
  {\mathcal H_{\rm eff.}} = -  J^*  \sum_{i,\sigma}  (c_{i,\sigma}^\dagger c_{i+1,\sigma} + {\rm H.c.}) + \omega^* \sum_i && \left(\beta^\dagger_i \beta_i + \frac{1}{2}\right)  \nonumber \\ 
  &+& 2 g_q  \sum_i  (\hat{n}_i-1) \left(\beta^\dagger_i \beta_i + \frac{1}{2} \right) - 4 \frac{g_q^2}{\omega} \sum_i  (\hat{n}_{i,\uparrow} - 1/2) (\hat{n}_{i,\downarrow} - 1/2) \left(\beta^\dagger_i \beta_i + \frac{1}{2} \right),
\end{eqnarray}
 with  $\omega^* = \tilde{\omega} + 2 \big(g_q + \frac{1}{2} \frac{g_q^2}{\omega} \big) = \omega -  \frac{g_q^2}{\omega}$.  Here, we have recast the Hamiltonian in terms of particle-hole symmetric electronic operators.  In this form, we see that the density-density interaction and the renormalized phonon terms are suppressed by a factor of $\omega/g_q \gg 1$ relative to the  particle-hole symmetric electronic density term responsible for disorder (third term), suggesting that in this limit the density term dominates the dynamics governed by ${\mathcal H_{\rm eff.}}$.

\subsection{Dynamics of observables in the squeezed frame}
The form of ${\mathcal H_{\rm eff.}}$ in Eq.\eqref{Eq:Heff.SI} (Eq.~\eqref{Eq:Heff.} of the main text) provides a simple view within which to understand the influence of the non-linear electron-phonon coupling on electronic properties.   However, in order to expose the intricate details of the dynamics on short and intermediate timescales, one must consider the  action of the transformation on the initial state and observables of interest. Consider an observable $\hat{O}$ measured in the original frame $\langle \hat{O} (t)\rangle = \bra{0} {\mathcal U}^\dagger(t) \hat{O} {\mathcal U}(t)  \ket{0}$, where $\ket{0} \equiv \ket{{\mathbf \Psi}}$ is the initial state and  ${\mathcal U}(t) = e^{-i{\mathcal H} t}$ is the time evolution operator for ${\mathcal H}$, Eqs.~\eqref{EqS1}-\eqref{EqS1details3} (Eq.~\eqref{Eq:QephHam} of the main text). We can obtain $\langle \hat{O} (t)\rangle$ equivalently in the squeezed frame defined by the transformation $U = e^S$ discussed above as $\langle \hat{O} (t)\rangle = \bra{\tilde{0}} \tilde{\mathcal U}^\dagger(t) \hat{\tilde{O}} \tilde{\mathcal U}(t)  \ket{\tilde{0}}$, where $\ket{\tilde{0}} = e^S \ket{0}$, $\tilde{\mathcal U}(t) = e^{-i\tilde{{\mathcal H}} t}$ and $\hat{\tilde{O}} = e^S \hat{O} e^{-S}$ are now in the rotated frame.   Evaluating $\langle \hat{O} (t)\rangle$  in this fashion comes with advantages. The squeezed Hamiltonian conserves  the squeezed phonon number, affording an analysis of the dynamics in terms of these constants of motion. In fact, we can make use of the approximations derived above and systematically consider the dynamics under the action of ${\mathcal H_{\rm eff.}}$ instead of $\tilde{{\mathcal H}}$ in the limit $g_q \ll \omega$ by making use of the following. We first write $\langle \hat{O} (t)\rangle = \bra{\tilde{0}} \tilde{\mathcal U}^\dagger(t) \hat{\tilde{O}} \tilde{\mathcal U}(t)  \ket{\tilde{0}} =  \bra{\tilde{0}} \mathrm{G}^\dagger(t) {\mathcal U_{\rm eff.}}^\dagger(t) \hat{\tilde{O}} {\mathcal U_{\rm eff.}}(t) \mathrm{G}(t) \ket{\tilde{0}}.$ Here ${\mathcal U}_{\rm eff.} = e^{-i{\mathcal H_{\rm eff.}}t}$ and $\mathrm{G}(t) = {\mathcal U}^\dagger_{\rm eff.} (t) \tilde{\mathcal U}(t)$, for which we can derive an integral equation of motion: $\mathrm{G}(t) = 1 - i \int_0^t dt' e^{i{\mathcal H_{\rm eff.}}t'} (\tilde{\mathcal H} - {\mathcal H_{\rm eff.}})e^{-i{\mathcal H_{\rm eff.}}t'} \mathrm{G}(t') $. We expand this expression for $\mathrm{G}(t)$ in linear response, retaining terms of $\mathcal{O}\{g_q/\omega\}$.  As discussed in the previous subsection, ${\mathcal H_{\rm eff.}}$ is exact to $\mathcal{O}\{g_q/\omega\}$.  Thus, the leading-order terms in $\tilde{\mathcal H} - {\mathcal H_{\rm eff.}}$ are already $\mathcal{O}\{(g_q/\omega)^2\}$.  We see that
$$\langle \hat{O} (t)\rangle  =  \bra{\tilde{0}} {\mathcal U_{\rm eff.}}^\dagger(t) \hat{\tilde{O}} {\mathcal U_{\rm eff.}}(t) \ket{\tilde{0}}$$ to $\mathcal{O}\{g_q/\omega\}$, and we simply need to consider the transformed initial state and observable under the action of ${\mathcal H_{\rm eff.}}$ instead of $\tilde{\mathcal H}$. Next, note that the transformation itself is parametrized  by the ratio $g_q / \omega$ in $\zeta_i$, which allows us to simplify this  expression by expanding $e^S \approx 1 + S$.  Thus, the equal-time expectation value of $\hat{O}$ to $\mathcal{O}\{g_q/\omega\}$ in the squeezed frame is
\begin{eqnarray} \label{Eq:LinearResponseSI}
&&\langle \hat{O} (t)\rangle  = \bra{0} {\mathcal U}^\dagger_{\rm eff.}(t) \hat{O} {\mathcal U}_{\rm eff.}(t) \ket{0} + \bra{0} {\mathcal U}^\dagger_{\rm eff.} (t) \Gamma_{\hat{O}}  {\mathcal U}_{\rm eff.}(t)  \ket{0} + \bra{0} {\mathcal U}^\dagger_{\rm eff.}(t) \hat{O} {\mathcal U}_{\rm eff.}(t)  \ket{g} + \bra{g} {\mathcal U}^\dagger_{\rm eff.}(t) \hat{O} {\mathcal U}_{\rm eff.}(t) \ket{0},
\end{eqnarray}
where $\Gamma_{\hat{O}} = \comm{S}{\hat{O}}$, and $\ket{g} = S\ket{0}$.  This result obtained within a linear response-like treatment consistently incorporates  $\mathcal{O}\{g_q/\omega\}$ corrections in the time evolution.
 
We are interested in the time evolution of the expectation values of electronic operators  $\hat{O} = \hat{O}_{\mathrm{e}}$ that depend on the charge or spin density. Noting that charge and spin correlations are conserved under the squeezing transformation ($\Gamma_C, \Gamma_S = 0$) and that terms in the above expression connecting $\ket{0}$ and $\ket{g}$ for $\hat{O} = \hat{O}_{\mathrm{e}}$ vanish at $\mathcal{O}\{g_q/\omega\}$ for real $\alpha$ used throughout this work, we arrive at (see Eq.~\eqref{Eq:LinearResponse} of the main text)
$$\langle \hat{O} (t)\rangle  = \bra{0} {\mathcal U}^\dagger_{\rm eff.}(t) \hat{O} {\mathcal U}_{\rm eff.}(t) \ket{0}.$$
To $\mathcal{O}\{g_q/\omega\}$, the dynamics of charge and spin correlations can be understood within an effective model in which we simply time evolve the initial phonon coherent state (now in the squeezed basis) under the action of ${\mathcal H_{\rm eff.}}$.

As we show in the main text, this effective model captures in a qualitative and sometimes semi-quantitative manner the behavior found in the exact results obtained in the unrotated frame. This simple model, however, provides evidence that the exact dynamics of the initial phonon coherent state is dominated by physical behavior given by its time evolution with $\mathcal{H}_{\rm eff.}$, which results in an ensemble of trajectories of independent conserved squeezed phonon configurations, and because the initial state is a Poisson linear superposition over phonon number states, this can be viewed as exactly equivalent to the disorder-averaged dynamics of a random system quenched to Poisson-distributed disorder, as has been established for models with binary disorder~\cite{ASmith1,ASmith2}.

The utility of the effective model as a descriptor of the behavior of electronic correlations may in fact extend to long times, but we have no means of confirming this since we only have access to correlation functions in infinite systems for which simulations are limited to short times.  Below we will discuss a complementary approach within which to understand the dynamics of electronic correlations in the original untransformed frame, which lends support to the persistence of disorder to longer times.

\section{Supplementary discussion of dynamics: Decoherence and heating}
In this section, we discuss supplementary details of the dynamics pertaining to decoherence and its influence on electronic dynamics, and heating.  

\subsection{Dephasing phonon-induced dynamics}
Similar to the approach of Ref.~\cite{NonLinear4Kennes}, we attempt to understand the influence of phonon coherence on the electronic dynamics. The main result we find is that the onsite phonon reduced density matrix in the phonon-number basis rapidly relaxes from its initial coherent state to a predominantly diagonal matrix, the elements of which form a unimodal distribution very similar to the  Poisson distribution that describes the eigenvalues of the initial phonon density matrix. This persistence of the diagonal character of the initial state phonon reduced density matrix accompanied by its rapid dephasing means that the approximation of the phonon distribution as an incoherent average over Poisson-distributed occupation-number eigenstates is reasonable, and lends support to the idea that disordered electron dynamics is intimately related to the nature of the initial state being a Poisson-distributed linear combination over phonon-number states.

To access these effects in the dynamics in the original unsqueezed frame from the exact data, we devise an approximate semi-classical {\em in silico} approach, which, using the exact phonon reduced density matrix extracted from the simulations, reproduces qualitatively the flattening in charge correlations.   This {\em in silico} approach in which one extracts information from the exact behavior of the phonons in order to reproduce the qualitative features of the dynamics of electronic correlations is not exact, but serves as a perspective on the influence of phonon decoherence on the electrons, complementary to the results obtained within the effective model presented in the main text.

\begin{figure*}
\centering
 \includegraphics[width=\columnwidth]{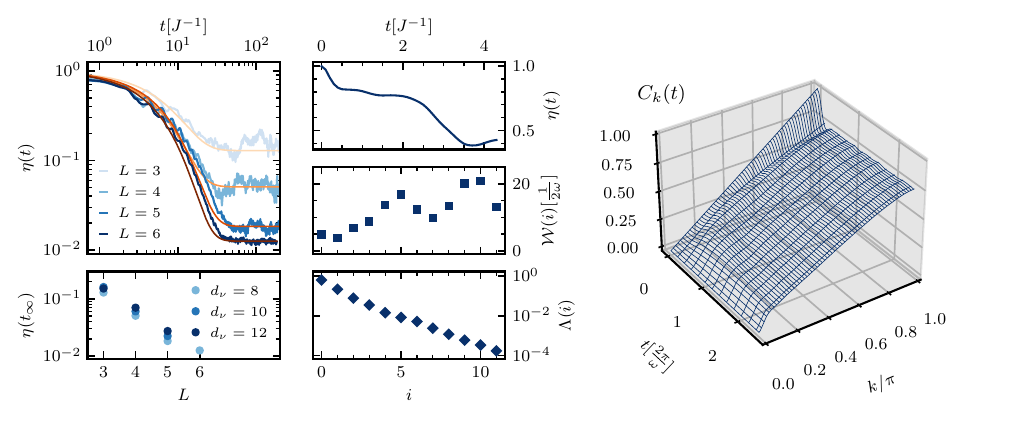}
 \caption{{\bf Dynamics of a metal subjected to a quadratic coupling, dephasing phonon-generated disorder.} {\em Left column:} Rapid loss of coherence in the onsite phonon reduced density matrix $\rho_{\rm ph}^{R}$ shown via analysis of $\eta(t) \equiv \sum_{\nu \neq \mu}|{\rho_{\rm ph}^{R}}_{\nu,\mu} (t)| / \sum_{\nu \neq \mu}|{\rho_{\rm ph}^{R}}_{\nu,\mu}(0)|$.  This is verified in Krylov propagation of systems of increasing size (left), and can be already observed on short timescales for infinite systems studied by iTEBD (center column, top). Thin lines in orange hues are fits of $\eta(t)$ to an exponential decay to a plateau (top). In the long-time limit, $\eta(t_{\infty})$ approaches increasingly vanishing values with larger system sizes (bottom). {\em Center column:} We use  the approach to diagonality of $\rho_{\rm ph}^{R}$ in iTEBD simulations (top) to invoke a semi-classical approximation in which we treat the phonons classically, as characterized by their reduced density matrix.  We extract a disorder potential from the coupled model for an exemplary time $t_q= \frac{2\pi}{\omega}$ via singular value decomposition of $\rho_{\rm ph}^{R}$, which we use to evaluate an effective classical disorder potential ${\mathcal W}(i)$ given by the expectation value of $\hat{X}^2$ in the singular vectors $i_{ {\mathcal S}(\rho_{\rm ph}^{R}) }$ (middle), and weighed by the probability distribution $\Lambda(i)$ given by the singular values (bottom). {\em Right column:}
We model the dynamics of the electrons quenched to the dephasing phonon potential given by ${\mathcal W}(i)$ weighted by the probability distribution $\Lambda(i)$, as prescribed by Eq.~\eqref{Eq:HAnderson.SI}. A free metal subjected to this disorder potential at initial time exhibits, after disorder averaging, a flattening charge correlator $C_k(t)$ with a  suppressed peak, qualitatively supporting the result of the fully coupled model observed in Fig.~\ref{fig:Flattening}.   We use $g_q = 0.25$ and $\omega = \pi/2$ in the simulations of the fully coupled model used in this figure.}
 \label{fig:Disorder2}
 \end{figure*}

We analyze the loss of coherence with time of the onsite oscillator reduced density matrix $\rho_{\rm ph}^{R}(t)$ in the phonon occupation-number basis.  We study the quantity $\eta(t) \equiv \sum_{\nu\neq \mu}|{\rho_{\rm ph}^{R}}_{\nu,\mu} (t)| / \sum_{\nu\neq \mu}|{\rho_{\rm ph}^{R}}_{\nu,\mu}(0)|$ ($\nu$ and $\mu$ are states of different phonon occupation number) as a measure of decoherence (Fig.~\ref{fig:Disorder2} left column, top and center column, top panels).  We find that $\eta(t)$ drops sharply from its initial value of unity corresponding to the pure initial phonon state to below 50\% at $t\sim4J^{-1}$ and to vanishingly small values in the long-time limit.  This implies that $\rho_{\rm ph}^{R}(t)$ evolves from its initial pure coherent state $\ket{\alpha}\bra{\alpha}$ to a mixed state that is predominantly diagonal in the phonon-number basis, signalling rapid dephasing of states with different phonon occupation number. The dephased configuration exhibits a unimodal distribution of diagonal matrix elements, which closely resembles the initial state Poisson distribution.  Our numerics reveals a strong  sensitivity of the electron dynamics to the approach of $\rho_{\rm ph}^{R}$ to diagonality, as also corroborated in finite-size systems in which we find the phonon coherence and electronic observables (e.g. energy density) both relax and approach the steady state on the same characteristic timescale $t\sim5J$ (not shown). This suggests that the diagonal matrix elements of the $\hat{X}^2$ operator can be  thought of as a slowly evolving classical dynamical onsite potential for the electrons (see also Fig.~\ref{fig:Pumpdynamics}, showing slow evolution with time of $\hat{X}^2$ and its correlation with charge at $t\sim5J$).  We may thus invoke a semi-classical approximation in which we neglect the rapidly decaying and small off-diagonal components of the $\hat{X}^2$ operator $\propto$ ${b^\dagger}^2$ and $b^2$, and model it as a classical diagonal variable that couples to the electron density in order to understand the influence of the non-linear coupling on the dynamics of electrons in terms of a dephasing phonon-generated disorder, which ultimately destroys the initial state quasi-long-ranged electronic density wave correlations.

To simulate this picture we consider an Anderson model for the dynamics of an initial state of a translation-invariant free-electron half-filled metal $\ket{\mathrm FS}$ with $k_{\rm F} = \pi/2$ evolved via a Hamiltonian that includes a static quenched onsite disorder potential extracted from the dephased phonon $\hat{X}^2$ obtained in exact simulations of the fully coupled model at intermediate times:
 \begin{eqnarray} \label{Eq:HAnderson.SI}
  {\mathcal H_{\rm Anderson}} = -  J \sum_{i,\sigma}  (c_{i,\sigma}^\dagger c_{i+1,\sigma} + {\rm H.c.}) +  \sum_i  \mathcal{E}_i \hat{n}_i
\end{eqnarray}
with $\mathcal{E}_i$ drawn from a classical disorder potential ${\mathcal W}(i)$ given by the expectation value of $\hat{X}^2$ in the the singular vectors $i \equiv i_{ {\mathcal S}(\rho_{\rm ph}^{R})}$ of the phonon reduced density matrix $\rho_{\rm ph}^{R}$ (i.e., $\mathcal{E}_i \in {\mathcal W}(i) = \bra{i_{ {\mathcal S}(\rho_{\rm ph}^{R})}} \hat{X}^2 \ket{i_{ {\mathcal S}(\rho_{\rm ph}^{R})}}$) with weights specified by the probability distribution $\Lambda(i)$ of singular values of $\rho_{\rm ph}^{R}$ over singular vectors $i$,  see Fig.~\ref{fig:Disorder2}, center column. We find that the momentum-resolved charge dynamics exhibits a rapid flattening  (Fig.~\ref{fig:Disorder2}, right column), bolstering the dephasing phonon-induced disorder picture of electron dynamics in the pumped metal.

\subsection{Estimates of electron heating and phonon relaxation}
An issue that arises naturally in the context of pump-probe experiments pertains to electronic heating due to phonon relaxation in the dynamics.  Here the initial pump creates an excited phonon state that couples to the electrons, and eventually relaxes by exchanging energy with the electronic and electron-phonon subsectors.  Transfer of energy to the electrons may ultimately destabilize transient phases that could have emerged outside of equilibrium. This proceeds in one of the following ways. Either the system, while still far from global thermal equilibrium, evolves to a different out-of-equilibrium state in which the electronic subsystem heats up to an effective temperature larger than the coherence temperature of the emergent phase, or the system eventually reaches true (global) thermal equilibrium at which point out-of-equilibrium behavior ceases to exist and the system becomes fully characterized by thermal distribution functions. In what follows we provide an analysis, based on numerics of finite size systems, of the  asymptotic long-time behavior of the energy redistribution amongst the system subsectors as a function of pump fluence and phonon adiabaticity, and  of the asymptotic expectation values of local electronic and phononic observables compared against their thermal expectation values which we use as proxy for the physical temperature of the electronic and phononic subsystems.

\subsubsection{Energetics as a function of pump fluence and phonon adiabaticity}
\begin{figure*}
\centering
 \includegraphics[width=\columnwidth]{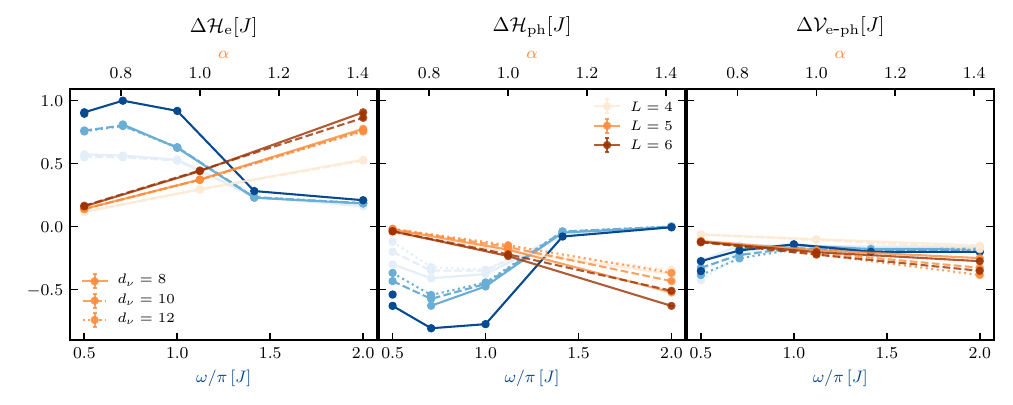}
 \caption{{\bf Net asymptotic change in electronic (left), phononic (center) and electron-phonon (right) energy densities as a function of $\alpha$, which sets the pump fluence (top horizontal axis), and  $\omega$, the phonon frequency (bottom horizontal axis).} Results are obtained from exact Krylov propagation of small systems $L=4-6$ with phonon local Hilbert space dimensions $d_{\nu} =8,10,12$ ($L=6$ is restricted to $d_{\nu} =8$) to asymptotically long times for the largest electron-phonon coupling $g_q = 0.25$.
 }
 \label{figS2}
 \end{figure*}

To understand the tendency for electronic heating and phononic relexation as a function of the pump excitation strength $\alpha$ and the adiabaticity regime set by $\omega$, we study the asymptotic net change in energy density of the electronic, phononic and electron-phonon subsectors in the long-time state obtained in finite-size simulations in Fig.~\ref{figS2}.  The results of Fig.~\ref{figS2} can be summarizes as follows.
\begin{itemize}[leftmargin=*]
    \item {\em Trend of energetics with increasing $\alpha$}: \\
        Electron heating increases with $\alpha$. Relaxation of the electron-phonon interaction energy increases with $\alpha$, and is non-vanishing even at smaller $\alpha$ (this is expected, since the interaction should have a stabilizing contribution). Phonon relaxation vanishes at the smallest $\alpha$ studied.   For even smaller $\alpha$, phonon heating becomes possible since the initial phonon state approaches  the phonon vacuum state as $\alpha \rightarrow 0$. 
    \item {\em Trend of energetics with increasing $\omega$}: \\
        Electronic heating and phonon relaxation exhibit non-monotonic behavior with $\omega$ with large changes in the interval $\omega \in [\pi/\sqrt{2}, \pi]J$. Dependence of these quantities on the system size decreases for larger $\omega$. In this limit, electron heating and phonon relaxation decrease with $\omega$ (the latter becomes basically negligible at the largest $\omega$), and the electron-phonon interaction energy stabilizes (plateaus) at large $\omega$. The change in the interaction energy is smallest for intermediate values of $\omega$ in contrast to the behaviour of the electron and phonon energies.
\end{itemize}
These results establish that in the adiabatic limit (small $\omega$) and for modest pump fluence ($\alpha = \sqrt{2}$), the long-time state exhibits non-vanishing electron-phonon correlations, accompanied by a net increase in the electronic energy.

\subsubsection{Long-time state electronic and phononic distribution functions}

Given that observables considered in finite-size simulations reach long-time plateaus with fluctuations that are consistent with behavior suggestive of equilibration, it is pertinent to ask whether the long-time steady state is thermal (in the Eigenstate thermalization hypothesis (ETH)~\cite{ETH} sense), i.e., if local observables have an expectation value consistent with their thermal expectation value at a temperature $T$ that corresponds to the energy density of the initial state. (Note that absence of thermalization in the small system sizes accessible in exact diagonalization (and propagation) is not  necessarily guaranteed to hold in the thermodynamic limit.) Here, we use  momentum-resolved electronic occupations $\langle \hat{n}_k \rangle$ and onsite phonon populations in the occupation number basis $\langle \hat{n}_{\nu} \rangle$ of the long-time state to judge whether it can be approximately considered to resemble a thermal state.  We contrast these against thermal expectation values which we obtain by computing the full Hamiltonian spectrum using exact diagonalization (ED) of finite-size systems.

\begin{figure*}
\centering
 \includegraphics[width=\columnwidth]{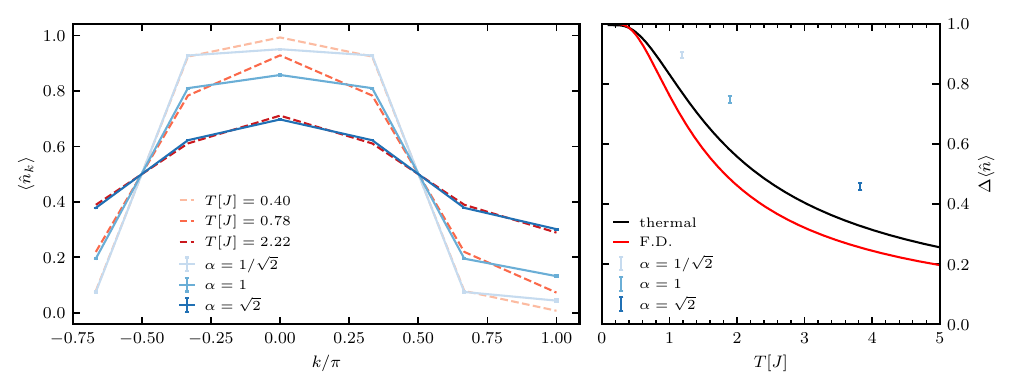}
 \caption{{\bf Electron distribution functions for different initial pump strength $\alpha$.}  {\em Left panel:} $\langle \hat{n}(k)\rangle$ of the long-time state from exact Krylov propagation of systems with  $L=6$ and $d_{\nu}=8$ (solid lines) and their fits to Fermi-Dirac distributions with temperature $T$ determined from the fit (dashed lines).  {\em Right panel:}  $\Delta \langle \hat{n} \rangle \equiv \langle \hat{n}(k=0) \rangle -\langle \hat{n}(k=\pi)\rangle$ for an $L=4$ ($d_{\nu} = 8$) system (this fully characterizes $\langle\hat{n}(k)\rangle$ on an $L=4$ system due to symmetry/conservation laws)  in the long-time state obtained from exact Krylov propagation with $T$ determined according to the initial state energy density (blue line error-bar markers; error-bars shows standard deviation of temporal fluctuations). We contrast  this against the thermal expectation value of $\Delta \langle \hat{n} \rangle$ at a given $T$ obtained from ED (solid black line) and  against the value of $\Delta \langle \hat{n} \rangle$ determined from a Fermi-Dirac distribution at a given $T$ (solid red line) of an $L=4$ system. All results are for $g_q=0.25$ and $\omega=\pi/2$.
 }
 \label{figS3}
 \end{figure*}

In Fig.~\ref{figS3} we study $\langle \hat{n}_k \rangle$ in the long-time state.  The exact electronic distribution function, at least at large $\alpha$, resembles a Fermi-Dirac distribution (Fig.~\ref{figS3}, left panel), however at a significantly lower $T$ than the physical temperature (Fig.~\ref{figS3}, right panel).  We also find that the long-time steady steady state for all $\alpha$ exhibits a more strongly peaked  distribution relative to its thermal counterpart obtained at $T$ that corresponds to the initial state energy density (Fig.~\ref{figS3}, right panel). Note that we could not find good fits  to Fermi-Dirac distributions  for long-time states obtained for small $\alpha$ values. These observations suggest that the electronic subsystem of the long-time state deviates from a thermal distribution.

\begin{figure*}
\centering
 \includegraphics[width=\columnwidth]{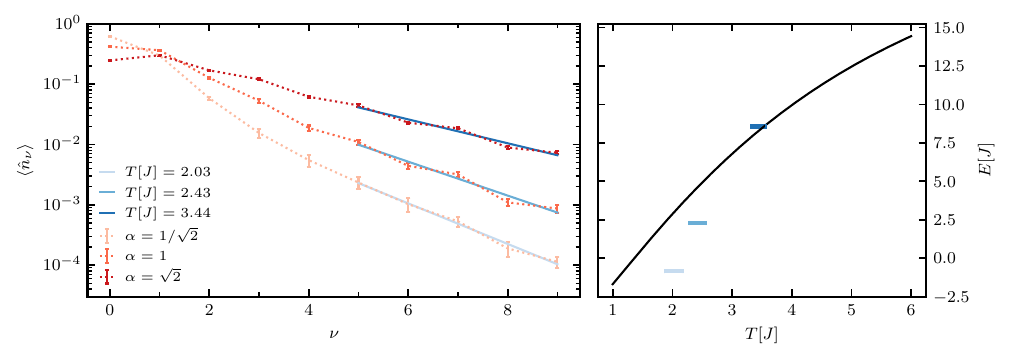}
 \caption{{\bf Phonon distribution functions for different initial pump strength $\alpha$.} {\em Left panel:} 
 $\langle \hat{n}_{\nu}\rangle$ of the long-time state from exact Krylov propagation of systems with  $L=4$ and $d_{\nu}=10$ (dotted lines) and the fits of their exponentially decaying tails to Maxwell-Boltzmann distributions with temperature $T$ determined from the fit (solid lines). {\em Right panel:}. Energy density of the initial state for a given $\alpha$ versus $T$ obtained from the fit of the tail of $\langle \hat{n}_{\nu}\rangle$ for the same $\alpha$ (blue horizontal lines whose length indicates temporal fluctuations obtained as standard deviation of the tail fits over different times in the long-time limit). This is compared to the thermal $E$ as a function of $T$ (solid black line) obtained from the full spectrum of the Hamiltonian computed in ED for $L= 4$, $d_{\nu} =10$. All results are for $g_q=0.25$ and $\omega=\pi/2$.
 }
 \label{figS4}
 \end{figure*}

In Fig.~\ref{figS4} we compute $\langle \hat{n}_{\nu} \rangle$ in the long-time state. The Poisson-like long-time phonon state found in finite systems exhibits a maximum in occupation numbers $\nu$ and therefore does not fit a thermal distribution. Of course, in a strongly coupled electron-phonon state a thermal phonon distribution is not expected.  However, the high-energy tail should still decay in a manner controlled by the equilibrium temperature  if the system has approached local equilibrium, and a fit of the  exponentially decaying phonon occupation tail to a Maxwell-Boltzmann distribution yields an effective temperature. Carrying out this analysis for the long-time state obtained from exact Krylov propagation, we find that the temperatures extracted from the phonon tail (Fig.~\ref{figS4}, left panel) overestimates the physical temperature (Fig.~\ref{figS4}, right panel), except at the largest $\alpha$. This analysis suggests that the long-time state of the system does not resemble a thermal state. Note that due to the underlying assumption regarding the phonon tail this constitutes less direct evidence of lack of equilibration than the comparison of the electronic distributions to thermal ones discussed above.

The evidence presented above indicates that, at least within the limited system sizes available to exact diagonalization and propagation, the system approaches a non-thermal long-time steady state. Drawing firm conclusions about thermalization from such small system sizes without proper finite-size scaling analysis (the latter being inaccessible to exact numerics) is of course not possible. A more thorough analysis of the existence or absence of thermalization and the associated timescales is left to future work.

\section{Comparison with the linearly coupled Holstein model}\label{QvHapp}
In this section we detail the methods we use to decide an appropriate value of the Holstein coupling to compare to a given value of the quadratic coupling.

The Holstein model with electron-phonon coupling $g_H (\hat{n}_i - 1) (b_i^\dagger + b_i)$ can be characterized via the dimensionless coupling $\lambda_H = \frac{g_H^2}{2 \omega J}$, the ratio of the ground-state energy in the atomic limit $J=0$ to that in the free electron limit $g_H = 0$. To compare the Holstein and quadratic models one must find the $\lambda_H$ most comparable to a given quadratic coupling $g_q$. We consider the two following approaches to estimate measures of equivalence of coupling strengths:
\begin{enumerate}[leftmargin=*,label=\alph*.]
    \item {\em Coupling strengths that give the same double occupancy in the static equilibrium limit}:\\
        We find for $\omega = \pi/2$,  $g_q = 0.25$ and $g_H = 0.29$ ($\lambda_H \approx 0.027$) yield the same double occupancy in the ground state of a half-filled chain. 
    \item {\em Coupling strengths that give the same effective electron-electron interaction obtained from a disentangling transformation}:\\
        The Lang-Firsov transformation~\cite{LangFirsov} demonstrates that Holstein phonons mediate an effective electron-electron attraction $U_H = -2 \frac{g_H^2}{\omega} \equiv -4 \lambda_H J$.  The squeezing transformation derived above demonstrates that quadratic phonons mediate an effective electron-electron attraction $U_q = -4 \frac{g_q^2}{\omega} \Big(\hat{n}_{{\rm B}_i}  + 1/2\Big)$ (recall $\hat{n}_{{\rm B}_i} = \beta_i^\dagger \beta_i$), see Eq.~\eqref{EqSbeta}.  The two models yield the same $U$ when $U_H = U_q$, leading to the condition:
        \begin{eqnarray}
        \lambda_H = \frac{g_q^2}{\omega J} \Big(\langle \hat{n}_{\rm B} \rangle + 1/2\Big),
        \end{eqnarray}
        where we replaced the phonon number operator by its average over the phonon distribution $\langle \hat{n}_{\rm B} \rangle$.  Since the radiation field creates a coherent state with amplitude $\alpha$, we take an estimate of $\langle \hat{n}_{\rm B} \rangle = \alpha^2$ the mean boson number to find $\lambda_H$ to be used to compare against a given $g_q$. We thus judge for $\alpha=\sqrt{2}$ and $\omega = \pi/2$ $\lambda_H \approx 0.1$ to be equivalent to $g_q = 0.25$ in the sense that it leads to an effective electron-electron interaction approximately equal to that obtained from the {\em pumped} quadratic model (as analyzed within the squeezing transformation).
\end{enumerate}
To summarize, we employ two methods to estimate a value of $\lambda_H$ to compare to a given value of $g_q$. One approach assumes the two models are comparable when they yield the same double occupancy in the static ground-state limit, the other compares the undriven Holstein model to the driven quadratic model, making use of analytical results.  We can conceptually use these two values of $\lambda_H$ as approximate lower and upper bounds for comparison against a given value of $g_q$.

\section{Details of numerical methods}
In this section we detail the numerical methods and employed convergence parameters used in the simulations of the non-linear electron-phonon model and of the effective  model.

\subsection{Details of simulations of the non-linear electron-phonon model}

\begin{figure*}
\centering
\includegraphics[width=0.925\columnwidth]{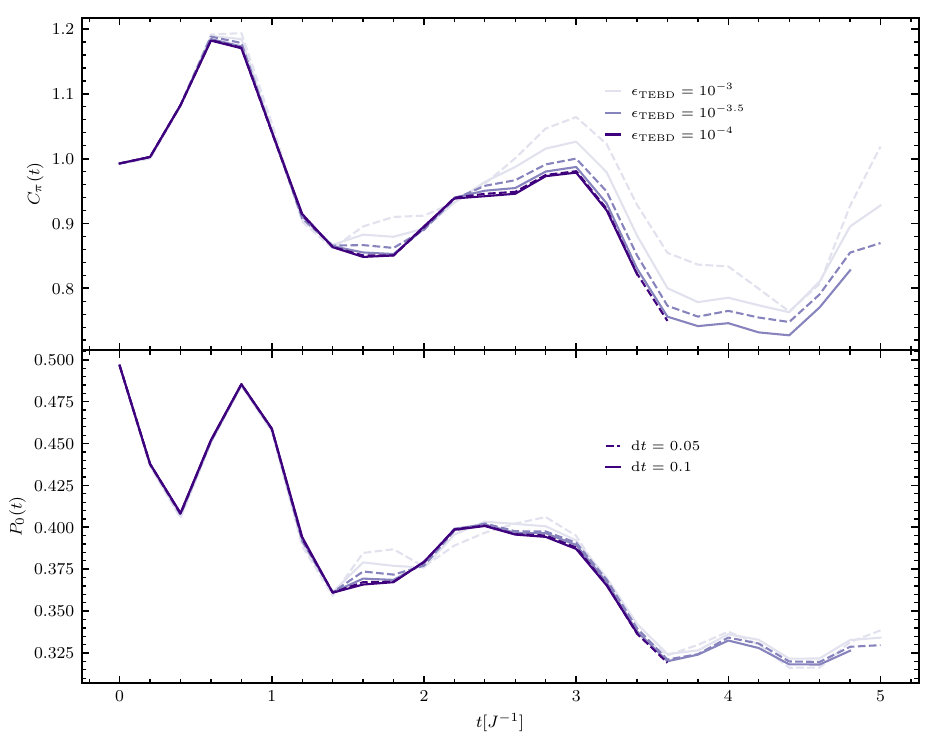}
\let\nobreakspace\relax
\caption{{\bf Convergence of time-evolved charge $C_k(t)$ and pairing $P_k(t)$ correlations with respect to truncation error $\epsilon_{\rm TEBD}$ and time-step ${\rm d}t$ used in iTEBD simulations.} We use $g_q=0.25$ and $\omega = \pi/2$ here, which enables the assessment of convergence for the strongest coupling and smallest phonon frequency considered. We observe satisfactory convergence for $\epsilon_{\rm TEBD} = 10^{-3.5}$ and ${\rm d}t=0.1$ on the accessible timescales.
}
 \label{figS5}
\end{figure*}

\begin{figure*}
\centering
\includegraphics[width=0.925\columnwidth]{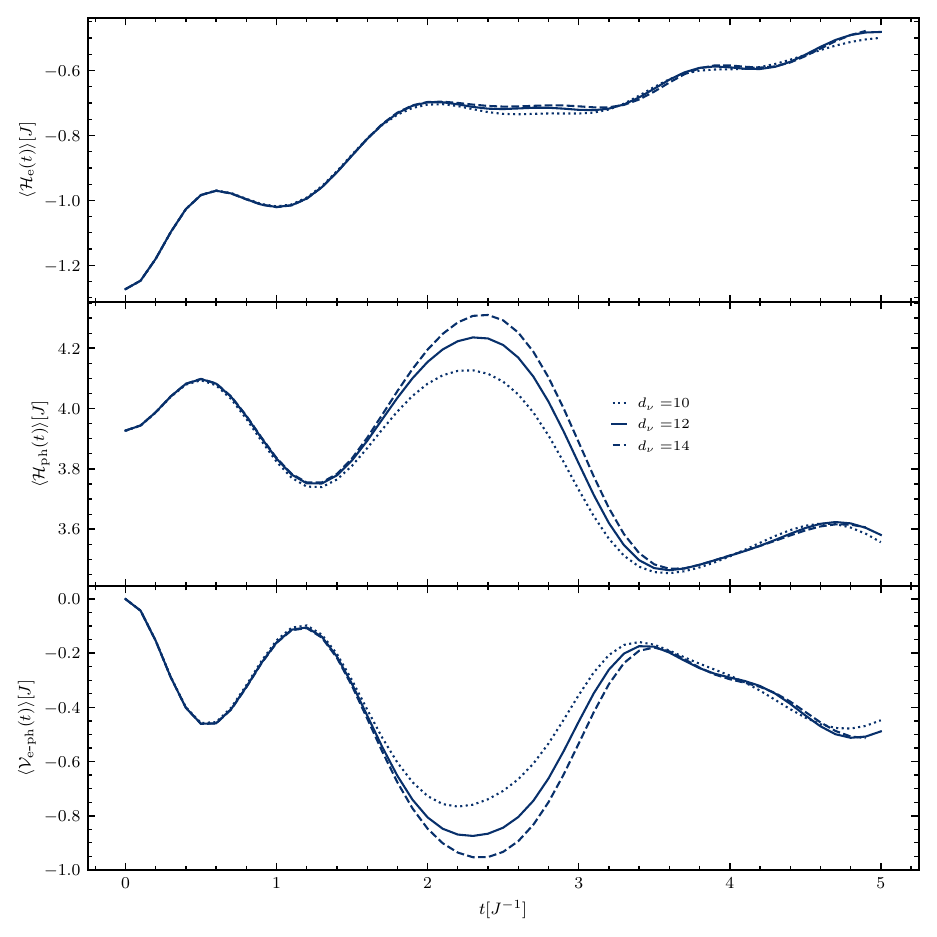}
\let\nobreakspace\relax
\caption{{\bf Convergence of time-evolved energy densities $\langle \mathcal{H}_{\rm e} (t) \rangle$, $\langle  \mathcal{H}_{\rm ph} (t) \rangle$ and
$\langle {\mathcal V}_{\rm e\mbox{-}ph}(t) \rangle$ with respect to the local phonon Hilbert space dimension $d_{\nu}$ used in iTEBD simulations.} We use $\epsilon_{\rm TEBD} = 10^{-3.5}$ in the simulation performed here for $g_q=0.25$ and $\omega = \pi/2$. We find that $d_{\nu} =12$ suffices to achieve convergence within a reasonable bound at all accessible times.}
 \label{figS6}
\end{figure*}

We simulate the time evolution of the initial state $\ket{0}\equiv \ket{\mathbf{\Psi}}$ under the action of the Hamiltonian of the non-linear electron-phonon model Eqs.~\eqref{EqS1}-\eqref{EqS1details3} (Eq.~\eqref{Eq:QephHam} of the main text) to intermediate timescales in infinite systems using iTEBD, and to long timescales in small systems using direct Krylov subspace methods.

\subsubsection{Details of \lowercase{i}TEBD simulations.}\label{TEBDapp}
The quadratic electron-phonon model connects a phonon state of occupancy $\nu$ only to states with $\nu\prime  = \nu \pm 2$. These processes conserve phonon parity. We take advantage of this symmetry and parallelize most simulations over even and odd phonon parity subsectors employing up to $d_{\nu} = 12$ states, see discussion below. We use a fourth-order trotterization scheme for the iTEBD time evolution with time-steps ${\rm d}t$.  After each time-step, we truncate the Schmidt values of a two-site unit cell state embedded in an infinite system; the discarded Schmidt values squared $\epsilon_{\rm TEBD}$ denotes the error due to truncation.  We ensure that the bond dimension $\chi$ of the time-evolved state after each time-step does not saturate an upper bound we set, which we take to be,  for the data points we study, in the range of $3000-5000$. We converge our results with respect to both ${\rm d}t$ and $\epsilon_{\rm TEBD}$, as we explain below.

\paragraph{Convergence with respect to ${\rm d}t$ and $\epsilon_{\rm TEBD}$.}

Errors due to ${\rm d}t$ compete with those due to $\epsilon_{\rm TEBD}$.  A sufficiently small ${\rm d}t$  ensures negligible Trotter error. At the same time, however, it results in more frequent incidents of truncation of the Schmidt values, each of an amount $\sqrt{\epsilon_{\rm TEBD}}$, thus leading to overall greater Schmidt truncation in order to access a specific desired final time $t_f$.  A sufficiently small $\epsilon_{\rm TEBD}$ would eliminate Schmidt errors to within a desirable accuracy, but instead  leads to faster growth of entanglement, which scales exponentially in time, and this limits the accessible $t_f$.  To ensure accurate results one needs to converge results with respect to the competing effects due to ${\rm d}t$ and $\epsilon_{\rm TEBD}$, finding an optimal compromise of a sufficiently small (but not too small) ${\rm d}t$ to eliminate Trotter error given a reasonably small  $\epsilon_{\rm TEBD}$ to ensure minimal error due to Schmidt truncation. In Fig.~\ref{figS5}, we demonstrate convergence for two quantities $P_k(t)$ and $C_k(t)$.  The same choices of ${\rm d}t$ and $\epsilon_{\rm TEBD}$ allows convergence of all other quantities considered in this work to the same standard or better.  This allows us to approach $t_f \sim 5 J^{-1}$.

\paragraph{Convergence with respect to $d_{\nu}$.}
We converge results for electronic and phononic observables with respect to the phonon Hilbert space dimension $d_{\nu}$ within a reasonable accuracy of a few percent. Fig.~\ref{figS6} shows satisfactory convergence of representative quantities for $d_{\nu} = 12$, which we use to obtain the data presented in the main text.

\subsubsection{Details of propagation using direct Krylov subspace methods}\label{Krylovapp}
 
We perform exact time evolution via direct Krylov space methods for system sizes $L=3-6$ with a twisted boundary condition: $e^{i(\pi/2)L}$, employing a  parallelization with respect to the local bosonic parity sectors. For small system sizes, convergence with respect to the local bosonic Hilbert space dimension can be achieved, while for $L=6$ we are restricted to a truncated bosonic Hilbert space dimension $d_{\nu} = 8, 10$.

\subsection{Details of simulations of the effective model}
We simulate the time evolution of the initial state $\ket{0}$ in the squeezed basis under the action of $\mathcal{H}_{\rm eff.}$ using iTEBD, employing $d_{\nu} = 12$ phonon states to accurately represent the initial coherent state.  We use a fourth-order trotterization scheme for the iTEBD time evolution with time-steps ${\rm d}t$.  After each time-step, we truncate the Schmidt values of a two-site unit cell state embedded in an infinite system. We ensure that the bond dimension $\chi$ of the time-evolved state after each time-step does not saturate an upper bound of $5000$. We converge our results with respect to both ${\rm d}t$ and $\epsilon_{\rm TEBD}$, finding that ${\rm d}t = 0.1$ and $\epsilon_{\rm TEBD} = 10^{-3.5}$ provide satisfactory convergence and access to timescales $t\sim 5J^{-1}$ for the largest coupling ($g_q = 0.25$) and smallest phonon frequency ($\omega = \pi/2$) considered.

\end{document}